\documentclass[11pt]{article}
\usepackage{textcomp}
\usepackage{amsmath}

\usepackage[breaklinks,pdfborderstyle={/S/U/W 1},%
  linkbordercolor={1 0 0},citebordercolor={1 0 0}]{hyperref}
\usepackage{gefpaper}

\newtheorem{theorem}{Theorem}
\newtheorem{lemma}[theorem]{Lemma}

\newcommand*{\set}[1]{\{#1\}} 
  \newcommand*{\st}{\mid} 
\newcommand*{\flor}[1]{\lfloor{#1}\rfloor} 
\newcommand{\ket}[1]{|{#1}\rangle} 
\newcommand{\braket}[2]{\langle{#1|#2}\rangle} 
\newcommand*{\quo}[1]{``#1''} 
\newcommand*{\mathquo}[1]{\mbox{``}#1\mbox{''}} 
\newcommand*{\xss}[1]{_{\scriptscriptstyle{#1}}} 
\newcommand*{\Uss}[1]{^{\scriptstyle{#1}}} 
\newcommand*{\str}{^{\ast}} 
\newcommand*{\rvs}{^{\scriptscriptstyle\rm R}} 
\newcommand*{\trs}{^{\scriptscriptstyle\rm T}} 
\newcommand*{\xth}{\textsuperscript{th}} 
\newcommand*{\xst}{\textsuperscript{st}} 

\newcommand*{\eps}{\ensuremath{\varepsilon}} 
\newcommand*{\blank}{\ensuremath{\#}} 
\newcommand*{\marker}{\ensuremath{\dashv}} 
\newcommand*{\dollar}{\ensuremath{\$}} 
\newcommand*{\ccent}{\ensuremath{\mathchoice
  {\mbox{\normalsize\textcent}}{\mbox{\normalsize\textcent}}%
  {\mbox{\scriptsize\textcent}}{\mbox{\tiny\textcent}}}} 
\providecommand*{\qeq}{\stackrel{{\scriptscriptstyle\mathrm{?}}}{=}} 
\newcommand*{\deltaq}{\delta_{\rm q}}
\newcommand*{\deltac}{\delta_{\rm c}}
\newcommand*{\bit}[2]{[#1]_{#2}}
\newcommand{\cmatrix}[2]{\mbox{\scriptsize
  $\left(\begin{array}{#1}#2\end{array}\right)$}} 

\newcommand*{\aml}{\hyperlink{hy:AM}{\ensuremath{\mathtt{AM}}}}
\newcommand*{\rei}{\hyperlink{hy:REI}{\ensuremath{\mathtt{REI}}}}
\newcommand*{\upower}{\hyperlink{hy:UPOWER}{\ensuremath{\mathtt{UPOWER}}}}
\newcommand*{\upowerp}{\hyperlink{hy:UPOWER}{\ensuremath{\mathtt{UPOWER+}}}}
\newcommand*{\power}{\hyperlink{hy:POWER}{\ensuremath{\mathtt{POWER}}}}
\newcommand*{\usquare}{\ensuremath{\mathtt{USQUARE}}}
\newcommand*{\clp}{\ensuremath{\mathcal{P}}}
\newcommand*{\clup}{\ensuremath{\mathcal{U}_{\scriptscriptstyle\mathcal{P}}}}
\newcommand*{\cla}{\ensuremath{\mathcal{A}}}
\newcommand*{\clap}{\ensuremath{\mathcal{A}'}}
\newcommand*{\cle}{\ensuremath{\mathcal{E}}}
\newcommand*{\si}{\ensuremath{s_{\scriptscriptstyle\rm I}}}
\newcommand*{\sa}{\ensuremath{s_{\scriptscriptstyle\rm A}}}
\newcommand*{\sr}{\ensuremath{s_{\scriptscriptstyle\rm R}}}
\newcommand*{\qi}{\ensuremath{q_{\scriptscriptstyle\rm I}}}

\newcommand*{\iit}[1]{\emph{\/(#1)\/}} 
\newcommand*{\lbr}{\linebreak[0]} 

\newcommand*{\bsection}[2]{\section[#1]{#1\label{#2}}\ignorespaces}
\newcommand*{\bsubsection}[2]{\subsection[#1]{#1\label{#2}}\ignorespaces}

\newcommand*{\btheorem}[1]{\begin{theorem}\label{#1}\ignorespaces}
\newcommand*{\etheorem}{\unskip\end{theorem}}

\newcommand*{\blemma}[1]{\begin{lemma}\label{#1}\ignorespaces}
\newcommand*{\elemma}{\unskip\end{lemma}}

\newcommand*{\bequations}{\begin{equation}\begin{array}{rcll}}
\newcommand*{\eequations}[1]{\end{array}\label{#1}\end{equation}\ignorespaces}

\newcommand*{\bfigure}{\unskip
  \begin{figure}[tp]\footnotesize\centering}
\newcommand{\efigure}[2]{\unskip\vspace{-0.8\baselineskip}%
  \caption[x]{#1\label{#2}}\end{figure}\ignorespaces}

\newcommand*{\btable}[2]{\unskip
  \begin{table}[tp]\footnotesize\centering
  \vspace{-0.8\baselineskip}\caption[x]{#1}\label{#2}\ignorespaces}
\newcommand{\etable}{\unskip\end{table}\ignorespaces}

\newcommand*{\bprogram}{\unskip\begin{tabbing}%
  \hspace*{\listparindent}\=\End \=\End \=\End \=\End \=\+\kill}
\newcommand*{\eprogram}{\unskip\end{tabbing}}
\newcommand*{\End}{\textbf{end} }
\newcommand*{\If}{\textbf{if} }
\newcommand*{\Then}{\textbf{then} }
\newcommand*{\Else}{\textbf{else} }
\newcommand*{\For}{\textbf{for} }
\newcommand*{\To}{\textbf{to} }
\newcommand*{\Do}{\textbf{do} }
\newcommand*{\Loop}{\textbf{loop} }
\newcommand*{\Exit}{\textbf{exit} }
\newcommand*{\Run}{\textbf{run} }
\newcommand*{\Accept}{\textbf{accept} }
\newcommand*{\Reject}{\textbf{reject} }

\newcommand*{\bitemize}{\begin{itemize}}
\newcommand*{\eitemize}{\unskip\end{itemize}}

\newcommand*{\bdisplay}{\[\begin{array}{rcll}}
\newcommand*{\edisplay}{\end{array}\]}

\providecommand*{\qed}{\mbox{}\nolinebreak\hfill~\raisebox{0.77ex}[0ex]{\framebox[1ex][l]{}}}
\newenvironment{proof}{\noindent\emph{Proof\/}:\hspace{\labelsep}\ignorespaces}{\bigbreak}
\newcommand*{\bproof}{\begin{proof}}
\newcommand*{\eproof}{\qed\end{proof}}

\begin{document}
\title{New Results on the Minimum Amount of Useful Space%
  \footnotetext[0]{\hyperlink{hy:RY14}{A~preliminary version} of
    this work was presented at the
    18\textsuperscript{th}~International Conference on Developments
    in Language Theory (DLT~2014), August 5--8, 2014, Ekaterinburg,
    Russia [{\em Lect.\ Notes Comput.\ Sci.},\ 8633, pp.~315--26\@.
    Springer-Verlag, 2014]\protect\nocite{RY14}\@.}%
}%
\author{Zuzana Bedn\'{a}rov\'{a}\mbox{\footnotemark[1] } and
  Viliam Geffert%
    \thanks{Supported by the Slovak Grant Agency for Science under contract VEGA~1/0142/15\@.}\\
    Department of Computer Science, P.\,J.\,\v{S}af\'{a}rik University, Ko\v{s}ice, Slovakia\\
    {\small\sf \textnormal{\{}zuzana.bednarova,\,viliam.geffert\textnormal{\}}@upjs.sk}\\[1.0ex]
  Klaus Reinhardt\\
    Wilhelm-Schickard-Institut f\"{u}r Informatik, University of T\"{u}bingen, Germany\\
    and Institut f\"{u}r Informatik, University of Halle, Germany\\
    {\small\sf klaus.reinhardt@uni-tuebingen.de}\\[1.0ex]
  Abuzer Yakary{\i}lmaz%
    \thanks{Supported by CAPES with grant 88881.030338/2013-01 and ERC Advanced Grant MQC\@.}\\
    National Laboratory for Scientific Computing, Petr\'{o}polis, RJ, Brazil\\
    {\small\sf abuzer@lncc.br}%
}%
\date{August 04, 2015}%
\maketitle\thispagestyle{empty}
\begin{quotation}\small\noindent\textbf{Abstract.}\hspace{\labelsep}%
  We present several new results on minimal space requirements to
  recognize a nonregular language: (i)~realtime nondeterministic
  Turing machines can recognize a nonregular unary language within
  weak $\log\log n$ space, (ii)~$\log\log n$ is a tight space lower
  bound for accepting general nonregular languages on weak realtime
  pushdown automata, (iii)~there exist unary nonregular languages
  accepted by realtime alternating one-counter automata within weak
  $\log n$ space, (iv)~there exist nonregular languages accepted by
  two-way deterministic pushdown automata within strong $\log\log n$
  space, and, (v)~there exist unary nonregular languages accepted by
  two-way one-counter automata using quantum and classical states
  with middle $\log n$ space and bounded error.\\
\mbox{}\\
\textbf{Keywords:}\hspace{\labelsep}%
  pushdown automata; counter automata; nondeterminism; alternation;
  quantum computation; unary languages; nonregular languages.
\end{quotation}

\bsection{Introduction}{s:intro}
The minimum amount of useful \quo{resources} which are necessary for
a finite automaton to recognize a nonregular language is one of the
fundamental research directions. Many different \quo{resources} have
been introduced, e.g., access to the input tape (realtime, one-way,
or two-way), computational mode of the model (deterministic,
nondeterministic, alternating, probabilistic, or quantum), type of
the working memory (counter, pushdown store, or tape), etc.
Moreover, unary languages need a special attention, since they may
require resources that are different {}from those for languages
built over general (or binary) alphabets. We shall focus on the
\emph{minimum amount of useful space} and present some new results.

First, we show that \emph{realtime} nondeterministic Turing
machines~(NTMs) can recognize unary nonregular languages in weak
$O(\log\log n)$ space.%
\footnote{Throughout the paper, $\log x$ denotes the binary
  logarithm of~$x$, unless otherwise specified\@.}
Second, if the worktape is replaced by a \emph{pushdown
store}\,---\,which gives realtime nondeterministic pushdown
automata~(PDAs)\,---\,we obtain the same result on the binary
alphabet. Third, we show that their deterministic counterparts,
\emph{two-way deterministic} PDAs, recognize nonregular languages
with strong $O(\log\log n)$ space. These bounds are tight, matching
the lower bound for two-way alternating Turing machines
(ATMs)~\cite{Iw93}\@. In the unary case, we know that one-way
nondeterministic PDAs recognize regular languages only, but their
alternating counterparts can simulate any ATM that uses $\Theta(n)$
space~\cite{CKS81}\@. Such power does not seem to hold if we replace
the pushdown store with a counter. Fourth, we show that realtime
alternating \emph{one-counter} automata recognize some nonregular
unary languages with weak $O(\log n)$ space. (By \quo{space} we mean
the value of the counter, rather than the length of its binary
representation\@.) Here we also present a trade-off to
\emph{alternation depth}\@. Fifth, we show that a two-way
deterministic one-counter automaton (2DCA) with \emph{two qubits}
(2QCCA) can recognize a nonregular unary language by using
$O(\log n)$ space on its counter for accepted inputs. Without qubits
and with $o(n)$ space, 2DCAs recognize only regular unary
languages~\cite{DG82B}\@.

Our results are presented in Section~\ref{s:main}, with a discussion
of the known results. We also identify some new directions and
formulate a few open questions. The reader is assumed to be familiar
with the classical computational models and so we provide only the
definition for 2DCAs using a fixed-size quantum memory (in
Section~\ref{s:definitions})\@. The proofs are put in the remaining
sections. We refer the reader to~\cite{SayY15} for short and
to~\cite{NC10} for complete references on quantum computation.

\bsection{Our Results and New Directions}{s:main}%
\bsubsection{Deterministic, Nondeterministic, and Alternating Machines}{s:main:dna}
It is known that \iit{i}~no weak $o(\log\log n)$ space bounded
alternating two-way Turing machine (TM) can recognize a nonregular
language and \iit{ii}~there exists a unary nonregular language
recognized by a deterministic two-way TM in strong $O(\log\log n)$
space~\cite{AM75,Iw93,Sz94b}\@. For one-way TMs, the bounds are
given in Table~\ref{tab:lower-bounds}, taken {}from a recent paper by
Yakary{\i}lmaz and Say~\cite{YS13B}, in which it was shown that all
these bounds are tight for almost all \emph{realtime}~TMs. However,
for realtime nondeterministic and alternating TMs accepting
\emph{unary} nonregular languages, it was left open whether the
double logarithmic lower bounds are tight. We solve this problem
positively (bold entries in Table~\ref{tab:lower-bounds}), so now we
have a complete picture for~TMs\@:
\btable{Minimum space used by one-way TMs for recognizing nonregular
  languages\@.}{tab:lower-bounds}
  \mbox{}\\
  \begin{tabular}{|l|l|l|l|l|l|l|}\hline
    & \multicolumn{3}{c|}{General input alphabet}
      & \multicolumn{3}{c|}{Unary input alphabet}
      \\ \cline{2-7}
    & Strong & Middle & Weak & Strong & Middle & Weak \\ \hline
    Deterministic TM & $\log n$ & $\log n$ & $\log n$ & $\log n$ & $\log n$ & $\log n$
      \\ \hline
    Nondeterministic TM & $\log n$ & $\log n$ & $\log\log n$ & $\log n$ & $\log n$
      & $\boldsymbol{\log\log n}$ \\ \hline
    Alternating TM & $\log n$ & $\log\log n$ & $\log\log n$ & $\log n$ & $\log n$
      & $\boldsymbol{\log\log n}$ \\ \hline
  \end{tabular}
\etable

\btheorem{t:realtime-TM}
  There exists a unary nonregular language accepted by a realtime
  nondeterministic Turing machine having a single worktape in weak
  $O(\log\log n)$ space.
\etheorem
In our construction, we use a working alphabet with more than $2$
symbols (except for the blank symbol) and so it is still open
whether we can obtain the same result with a binary working
alphabet.

By using a TM with a restricted access to working tape, we obtain a
pushdown automaton~(PDA)\@. It is known that no weak $o(n)$ space
bounded one-way deterministic PDA can recognize a nonregular
language~\cite{Gab84} and that realtime deterministic PDAs can
recognize $\set{a^nb^n\st n\ge 0}$ in strong $O(n)$ space. For
one-way nondeterministic PDAs, a weak $O(\log n)$ space algorithm
was given for a nonregular language in~\cite{Re07}\@. We improve
this to weak $O(\log\log n)$ space. For this purpose, we introduce,
in Section~\ref{s:rtNPDA}, a new language called~\rei\ (used also in
Section~\ref{s:2DPDA})\@.

\btheorem{t:realtime-NPDA}
  Realtime nondeterministic PDAs can recognize nonregular language
  \rei\ with weak $O(\log\log n)$ space.
\etheorem

Since a pushdown automaton is a special case of the Turing machine,
this bound is tight also for any kind of alternating PDAs,
by~\cite{Iw93,Sz94b}\@. On the other hand, we do not know the tight
strong/middle space bounds for one-way/realtime nondeterministic and
alternating PDAs recognizing nonregular languages.

In the unary case, one-way nondeterministic PDAs cannot recognize
nonregular languages~\cite{GR62}\@. Realtime alternating one-counter
automata~(CAs), on the other hand, can recognize some unary
nonregular languages even in weak $O(\log n)$ space (counter
\hypertarget{hy:UPOWER}{}%
value)\@. Here we shall use the following two unary languages:
\bdisplay
  \upower = \set{a^{2^n}\st n\ge 0}
  &\mbox{ \ and \ }&
  \upowerp = \set{a^{2^n+4n-4}\st n\ge 3} \,.
\edisplay

\btheorem{t:realtime-A1CA}
  Realtime alternating CAs can recognize nonregular \upowerp\ in
  weak $O(\log n)$ space.
\etheorem
The tight space bounds for realtime/one-way alternating counter
automata recognizing nonregular unary/binary languages are still not
known.

In Section~\ref{s:realtime-A1CA}, we present a one-way algorithm for
\upower\ and then our realtime algorithm for \upowerp\@. Both
algorithms have a linear alternation depth on accepted inputs. In
Section~\ref{s:trade-off}, we consider the existence of a shorter
alternation depth and present a realtime algorithm for \upower\
(a~slightly modified version of the algorithm provided by
\v{D}uri\v{s}~\cite{Dur13A}) with alternation depth bounded by
$O(\log n)$, but it needs a linear counter. Moreover, we show that
if the counter is replaced by a pushdown store, we have only a
single alternation, using linear space.

In the case of two-way PDAs, we have tight bounds:

\btheorem{t:two-way-DPDA}
  Two-way deterministic PDAs can recognize \rei\ in strong
  $O(\log\log n)$ space.
\etheorem

In~\cite{DG82B}, it was shown that any unary language recognized by
a two-way deterministic PDA using $o(n)$ space is regular. Moreover,
two-way deterministic CAs can recognize nonregular unary \upower\
with $O(n)$ space. Therefore, linear space is a tight bound for both
two-way deterministic PDAs and CAs, but we do not know whether
nondeterministic or random choices can help for unary languages.

Another interesting direction is to identify the tight bounds for
one-way/realtime multi-counter/pushdown automata. Yakary{\i}lmaz and
Say~\cite{YS13B} showed that realtime deterministic automata with
$k$~counters can recognize some nonregular languages in middle
$O(n^{1/k})$ space, where $k>1$\@. The same result can be obtained
by bounded-error probabilistic one-counter automata, but the error
bound increases in~$k$\@.

\bsubsection{Probabilistic and Quantum Machines}{s:main:pq}
Clearly, probabilistic models are special cases of their quantum
counterparts. In the unbounded error case, realtime probabilistic
finite automata~(PFAs) can recognize unary nonregular
languages~\cite{Paz71}, so let us consider the bounded error case.
One-way PFAs with bounded-error recognize only regular
languages~\cite{Ra63}\@. Two-way PFAs can recognize some nonregular
languages but only with exponential expected
time~\cite{Fre81,DS90}\@. With an arbitrarily small (not constant)
space, two-way probabilistic TMs can recognize nonregular languages
in polynomial time~\cite{FK94}, but one-way probabilistic TMs do not
recognize nonregular languages in space
$o(\log\log n)$~\cite{Fre83,KF90}\@.

Two-way quantum finite automata~(QFAs), on the other hand, can
recognize some nonregular languages in polynomial
time~\cite{AW02}\@. If the input head is quantum, i.e., it can be in
a superposition of several places on the input tape, then one-way
QFAs can recognize some nonregular languages in linear
time \cite{KW97,AI99,YS09B,Yak12C}\@. But, it is not known whether
two-way QFAs can recognize any nonregular unary language with
bounded-error (which is not possible for 2PFAs~\cite{Kan91B})\@.

We shall show that two-way QFAs with a classical counter\,---\,or
2DCAs with a fixed-size quantum memory~(2QCCAs)\,---\,can recognize
a nonregular unary language with space smaller than required by the
deterministic 2DCAs.

\btheorem{t:quantum-logspace-counter}
  The unary nonregular language \upower\ can be recognized by a
  2QCCA with bounded-error, using middle $O(\log n)$ space on its
  counter.
\etheorem

One-way probabilistic PDAs cannot recognize nonregular unary
languages with bounded-error~\cite{KGF97} but the question is open
for their quantum counterpart. On the other hand, using middle
$O(\log n)$ space, realtime bounded-error probabilistic PDAs can
recognize the language
$\set{b_1ab_2\rvs ab_3ab_4\rvs a\cdots ab_{2k-1}ab_{2k}\rvs
  \st k>0}$,
where $\alpha\rvs$~denotes the reversal of a string~$\alpha$ and
$b_i$~the binary representation of~$i$\@. Currently, we do not know
any better  result and whether quantumness helps.

\bsection{Definitions}{s:definitions}
We use three different modes of space usage~\cite{Sz94b}\@:
\iit{i}~\emph{Strong space}~$s(n)$ refers to the space used by the
machine along all computation paths on all inputs of length~$n$,
\iit{ii}~\emph{middle space} to the space used along all computation
paths on accepted inputs, and \iit{iii}~\emph{weak space} to an
accepting path using minimum space.

A~\emph{one-way} machine model a restricted two-way variant never
moving the input head to the left. A~\emph{realtime} machine a
restricted one-way variant in which the input head can stay on the
same symbol only a fixed number of steps. (Actually, the realtime
machines presented in this paper never wait on the same symbol\@.)

A~two-way one-counter automaton with quantum and classical states
(2QCCA)~\cite{Yak13B} is a two-way one-counter automaton having a
constant-size quantum register. Without the counter, we obtain a
two-way finite automaton with quantum and classical states
(2QCFA)~\cite{AW02}\@. In the original definition, the automaton can
apply unitary and measurement operators to its quantum part. Here we
allow to apply a superoperator (see Figure~\ref{f:superoperators}),
a generalization of classical and unitary operators including
measurement. In general, this does not change the computational
power of 2QCFAs and 2QCCAs~\cite{AW02}\@. We only do not know
whether the original models are less powerful than the ones with
superoperators if we use only rational amplitudes. (All quantum
algorithms presented here use rational superoperators\@.)
\bfigure
  \fbox{\begin{minipage}{0.95\textwidth}%
    A~superoperator $\cle= \set{E_{1},\ldots,E_{k}}$ is composed of
    some operation elements $E_{i}$ satisfying
    \bdisplay
      \sum_{i=1}^{k} E_{i}^{\dagger} E_{i} &=& I,
    \edisplay
    where $k>0$ is a constant, and the indices are the measurement
    outcomes. When the superoperator is applied to the quantum
    register in a state~$\ket{\psi}$, i.e.,~$\cle(\ket{\psi})$, we
    obtain the measurement outcome $i\in \set{1,\ldots,k}$ with
    probability
    $p_{i}= \braket{\widetilde{\psi_{i}}}{\widetilde{\psi_{i}}}$,
    where~$\ket{\widetilde{\psi_{i}}}$, \emph{the unconditional
    state vector}, is calculated as
    $\ket{\widetilde{\psi}_{i}} = E_{i}\ket{\psi}$\@. Note that
    using unconditional state vector simplifies calculations in many
    cases. If the outcome~$i$ is observed with $p_{i}>0$, the new
    state of the system, which is obtained by
    normalizing~$\ket{\widetilde{\psi}_{i}}$, is given by
    $\ket{\psi_{i}}= \ket{\widetilde{\psi_{i}}}/\sqrt{p_{i}}$\@.
    Moreover, as a special operator, the quantum register can be
    initialized to a predefined quantum state. This initialization
    operator has only one outcome.%
  \end{minipage}}
  \\ \mbox{}
\efigure{The details of superoperators~\cite{Yak13C}\@.}{f:superoperators}

A~2QCCA is an $8$-tuple
\mbox{$\clp\!=\!(S,Q,\Sigma,\delta,\si,\sa,\sr,\qi)$}, where
$S$ denotes the set of classical states,
\mbox{$\si$,$\sa$,$\sr\!\in\!S$} (with \mbox{$\sa\!\ne\!\sr$}) the
initial, accepting, and rejecting states, $Q$~the set of quantum
states, \mbox{$\qi\!\in\!Q$} the initial quantum state, $\Sigma$~the
input alphabet (not containing \ccent,\dollar, the left and right
endmarkers), and $\delta$~the transition function composed of
$\deltaq$ and~$\deltac$ governing the quantum and classical parts,
respectively.

Such machine~\clp\ starts with the given input $w\in\Sigma\str$
enclosed in between \ccent\ and~\dollar, the input head placed
on~\ccent, in the state $(\,\ket{\qi},\si)$, and zero in the
counter. Now, if \clp\ is in a state $(\,\ket{\psi},s)$ with the
input head on a symbol~$a$ and with
\mbox{$\theta\in \set{\mathquo{\!=\!0}\!,\!\mathquo{\!\ne\!0}}$},
the next step consists of the following quantum and classical
transitions. First, $\deltaq(s,a,\theta)$ determines a superoperator
which is applied to the quantum register and some classical
outcome~$\tau$ is observed. The quantum part of the state is updated
to~$\ket{\psi_\tau}$\@. After that, if
$\deltac(s,a,\theta,\tau)=(s',d,c)$, the classical part of the state
changes to~$s'$ and the input head and the counter value are updated
with respect to $d\in \set{\leftarrow,\downarrow,\rightarrow}$ and
$c\in \set{-1,0,+1}$\@. \,\clp~accepts or rejects when it enters \sa\
or~\sr, respectively.

\bsection{Realtime Nondeterministic Turing Machine -- Theorem~\ref{t:realtime-TM}}{s:realtime-TM}
\hypertarget{hy:fn}{}%
The following function will play an important role in our
considerations:
\bdisplay
  f(n) &=& \mbox{the smallest positive integer not dividing $n$} \,.
\edisplay
We take $f(0)=+\infty$ (i.e., undefined): there is no positive
integer not dividing~$0$\@. We shall use the fact that $f(n)$ can be
written down with $O(\log\log n)$ bits~\cite{AM75,FL75,Sz94b}\@.
However, we shall need to be more precise about the constants hidden
in the \mbox{big-$O$} notation:%
\footnote{The upper bound on~$f(n)$ will be required not only to
  derive an upper bound for space, but also for tuning up some
  parameters, so that our algorithm will work correctly\@.}

\blemma{l:fn}
  $f(n)< 2\!\cdot\!\log n$, for each $n\ge 3$\@.
\elemma
\bproof
\iit{i}~Consider first the case of $n\ge 363$\@. For the given~$n$,
let $f(n)= 1\!+\!m$ be the smallest positive integer not
dividing~$n$\@. Consequently, each $k\in \set{1,\ldots,m}$
divides~$n$\@. Thus, $p^{\flor{\log_p m}}$ must divide~$n$ for each
prime~$p$, since $1\le p^{\flor{\log_p m}}\le m$\@. But then $n$ is
a common multiple of all values~$p^{\flor{\log_p m}}\!$\@. Second, for
primes satisfying $p\le m$, we have $\log_p m\ge 1$, and hence also
$\flor{\log_p m}\ge 1$\@. Combining this, we get:
\bdisplay
  n &\ge& \prod_{p\le m} p^{\flor{\log_p m}} =
    \prod_{p\le m} (p^{\flor{\log_p m}+\flor{\log_p m}})^{1/2} \ge
    \prod_{p\le m} (p^{1+\flor{\log_p m}})^{1/2} \\[1.0ex]
  &>& \prod_{p\le m} (p^{\log_p m})^{1/2} = \prod_{p\le m} m^{1/2} =
  m^{1/2\cdot\pi(m)} ,
\edisplay
where all products are taken over primes $p\le m$ and $\pi(m)$
denotes the total number of primes smaller than of equal to~$m$\@.
{}From~\cite{RS62}, we know that $\pi(x)> \frac{x}{\ln x}$, for each
real $x\ge 17$\@. Using this, we have the following two subcases:

If $m\ge 17$, we can bound~$n$ {}from below as follows:
\bdisplay
  n &>& m^{1/2\cdot\pi(m)} > m^{1/2\cdot m/\ln m} =
    e^{\ln m\cdot 1/2\cdot m/\ln m} = e^{1/2\cdot m} ,
\edisplay
and hence $\frac{m}{2}< \ln n$\@. Now, using
$1< 0.6\!\cdot\!\log 4$, \,$4< n$, and $2\!\cdot\!\ln 2< 1.4$, we
get
\bdisplay
  f(n) &=& 1+m < 0.6\!\cdot\!\log 4 + 2\!\cdot\!\ln n <
    0.6\!\cdot\!\log n + 2\!\cdot\!\ln 2\!\cdot\!\frac{\ln n}{\ln 2} \\
  &<& 0.6\!\cdot\!\log n + 1.4\!\cdot\!\log n = 2\!\cdot\!\log n \,.
\edisplay

Conversely, if $m< 17$, that is, if $m\le 16$, we can use the fact
that $\frac{17}{2}< \log 363\le \log n$\@. This gives:
\bdisplay
  f(n) &=& 1 + m \le 17 < 2\!\cdot\!\log n \,.
\edisplay

\iit{ii}~It only remains to prove the statement of the lemma for
$n< 363$\@. However, it is a routine task to compute the table of
values $\frac{f(n)}{\log n}$, for $n= 362,\ldots,3$, and verify that
each of these values is smaller than~$2$\@.
\eproof

\hypertarget{hy:AM}{}%
Consider now the following unary language:
\bdisplay
  \aml &=& \set{1^n\st
    \mbox{\hyperlink{hy:fn}{$f(n)$} is not equal to a power of $2$}} \,.
\edisplay
Historically, the \emph{complement} of~\aml\ was the first known
\emph{unary nonregular language} accepted with only $O(\log\log n)$
space, by a strongly bounded two-way deterministic Turing
machine~\cite{AM75}\@. Later, in~\cite{BMP95} (see
also~\cite{Me08}), it was shown that \aml~can be accepted by a
\emph{weakly bounded one-way} nondeterministic machine, with
$O(\log\log n)$ space again. The machine in~\cite{BMP95} is based on
the observation that, for each $n> 0$,
\bitemize
  \item $1^n\!\in\!\aml$ if and only if there exist two positive
    integers $k$ and~$i$ satisfying $2^i< k< 2^{i+1}\!$, such that
    $n\bmod k\ne 0$ \ and \ $n\bmod 2^i=0$\@.
  \item Moreover, for $1^n\!\in\!\aml$, the membership can be
    certified by taking $k=f(n)$ and $2^i= 2^{\flor{\log k}}\!$\@.
    This gives $2^i< k< 2\cdot\log n$, by Lemma~\ref{l:fn}\@.
\eitemize

This machine accepts~\aml\ as follows: \iit{i}~Guess some $k,2^i$
satisfying $2^i< k< 2^{i+1}\!$\@. \iit{ii}~Traversing across the
input, count $r\xss{1}= n\bmod k$ \ and \
$r\xss{2}= n\bmod 2^i\!$\@. That is, at each input tape position,
execute $r\xss{1}:= (r\xss{1}\!+\!1)\bmod k$ \ and \
$r\xss{2}:= (r\xss{2}\!+\!1)\bmod 2^i\!$\@. \iit{iii}~If the end of
the input is reached with $r\xss{1}\ne 0$ and $r\xss{2}= 0$, accept.

The values $k,2^i,\lbr r\xss{1},r\xss{2}$ are stored in binary in
four worktape tracks, \quo{one above another}$\!$, using some
$\ell$~worktape cells\@.%
\footnote{To present binary written numbers as usual, with the least
  significant bits on the right end, we use a worktape growing to
  the left, initially empty, with the right endmarker~\quo{\marker}
  and infinitely many blank symbols~\quo{\blank} to the left of it.
  (See also Figure~\ref{f:worktape}\@.) On the other hand, there is
  no endmarker at the end of the input: after reading the last input
  symbol, the machine stops immediately\@.}
Since $2^i,r\xss{1},r\xss{2}$ are all smaller than~$k$ and we can
choose $k=f(n)$, bounded by $2\!\cdot\!\log n$, an accepting
computation path using optimal amount of space works with
$\ell\le O(\log\log n)$ worktape cells\@.%
\footnote{Carefully implemented, we can increase and test
  $r\xss{1},r\xss{2}$ in their respective tracks simultaneously, by
  a single \emph{double-sweep} across the worktape (moving {}from
  the right worktape endmarker~\quo{\marker} to the first blank
  symbol~\quo{\blank}, followed by going back)\@. Since there are
  exactly $\ell$ nonblank symbols in between \quo{\blank}
  and~\quo{\marker}$\!$, we perform this with exactly
  $2\!\cdot\!(\ell\!+\!1)$ steps, per each input tape position. More
  implementation details will be presented for a more advanced
  realtime version\@.}

\bsubsection{A~Realtime Version -- Main Idea}{s:epsfree}
In what follows, suppose that $n> 0$ and $f(n)\ge 17$\@. (We shall
later see how to avoid this assumption\@.) But then $n$ must be a
common multiple of $\set{1,\ldots,16}$ different {}from zero, and
hence this assumption can be formulated as follows:
\bequations
  f(n)\ge 17  &\mbox{ \ and \ }&  n\ge 720720 \,.
\eequations{e:fnn}

Now, we would like to accept~\aml\ without stationary moves on the
input. First, take the machine with the sweeping worktape head,
discussed above. In this machine, we modify every single operation
so that we make the input head move one position forward. But then
the progress in the input head movement becomes
$2\!\cdot\!(\ell\!+\!1)$ times faster than the progress in modular
incrementing of $r\xss{1},r\xss{2}$ on the worktape. Recall that our
machine~\cla\ requires exactly $2\!\cdot\!(\ell\!+\!1)$ steps to
execute $r\xss{1}:= (r\xss{1}\!+\!1)\bmod k$,
\,$r\xss{2}:= (r\xss{2}\!+\!1)\bmod 2^i\!$, and to test whether
$r\xss{1}\ne 0$, \,$r\xss{2}= 0$, all this by one double-sweep, per
each input tape position.

However, we can increment faster, by executing
$r\xss{1}:= (r\xss{1}\!+\!\Delta)\bmod k$ and, simultaneously,
$r\xss{2}:= (r\xss{2}\!+\!\Delta)\bmod 2^i\!$, where $\Delta> 1$ is
a value stored in binary in a separate worktape track. The modified
machine~\cla\ still uses a single double-sweep across the worktape,
with exactly $2\!\cdot\!(\ell\!+\!1)$ steps, during which the input
head travels forward exactly $2\!\cdot\!(\ell\!+\!1)$ positions.
Thus, using $\Delta= 2\!\cdot\!(\ell\!+\!1)$, the progress in the
input head movement agrees with the progress in modular
incrementing. But then, for each input tape position~$r$ that is an
integer multiple of $2\!\cdot\!(\ell\!+\!1)$, the machine gets to
the position~$r$ with the corresponding worktape tracks containing
$r\xss{1} =r\bmod k$ \ and \ $r\xss{2} = r\bmod 2^i\!$\@. Moreover,
at each such input position, the machine \quo{knows} whether
$r\xss{1}\ne 0$ and $r\xss{2}= 0$, keeping this information in the
finite state control. Thus, if~$n$ is an integer multiple of
$2\!\cdot\!(\ell\!+\!1)$, we can correctly decide between acceptance
and rejection. (This also requires to initialize the worktape with
exactly $2\!\cdot\!(\ell\!+\!1)$ steps, assigning initially
$r\xss{1}:= (2\!\cdot\!(\ell\!+\!1))\bmod k$ \ and \
$r\xss{2}:= (2\!\cdot\!(\ell\!+\!1))\bmod 2^i$\@.)

But a problem arises if~$n$ is not an integer multiple of
$2\!\cdot\!(\ell\!+\!1)$\@. In this case, \cla~gets to the end of
the input at the moment when it is busy with computing in the middle
of the worktape. Thus, the respective worktape tracks for
$r\xss{1},r\xss{2}$ contain some intermediate data, {}from which
\cla~cannot quickly deduce whether $n\bmod k$ \ or \ $n\bmod 2^i$ is
equal to zero. This problem is resolved by tuning up the size of the
worktape more carefully. Namely, if $2\!\cdot\!(\ell\!+\!1)= 2^j\!$,
for some power of two satisfying $2^j\le 2^i\!$, the machine can get
to the end of the input while working in the middle of the worktape
only if~$n$ is not an integer multiple of~$2^j\!$\@. But then $n$ is
not an integer multiple of~$2^i\!$, and hence $n\bmod 2^i\ne 0$\@.
In this case, \cla~rejects.

Now we are ready to summarize all values, kept in six separate
worktape tracks, as well as requirements on these values. (This can
also be seen in Figure~\ref{f:worktape}\@.)
\bfigure
  \begin{tabular}[b]{@{}r@{\,\vline}%
      l@{\,\,\,}c@{\,}c@{}c@{}c@{\vline}%
      l@{\,\vline}l@{\ \ \ \ }l@{\ }l@{}}
    \multicolumn{2}{@{}r@{}}{\scriptsize $\ell\!-\!1\!\cdots\!i\!+\!1$}
      & \scriptsize $i$ &
      \scriptsize $i\!-\!1\!\cdots\!j\!+\!1$ & \scriptsize $j$ &
      \multicolumn{2}{@{}l@{}}{\scriptsize $j\!-\!1\cdots 0$} & &
      Track: & Granted: \\ \cline{1-7} \cline{9-10}
    & $\,\,0\ \cdots\ 0$ & $1$ & $0\!/\!1\!\cdots\!0\!/\!1$ &
      $0\!/\!1$ & $\,\,0\!/\!1\!\cdots\!0\!/\!1\,\,$ & & &
      $k$ & $2^i< k< 2^{i+1^{\phantom{X}}}$ \\ \cline{2-6}
    & $\,\,0\ \cdots\ 0$ & $1$ & $0\ \cdots\ 0$ & $0$ & $0\ \cdots\ 0$ & & &
      $2^i$ & $2^{i+1}\le 2^{\ell}$ \\ \cline{2-6}
    $\cdots\blank$ & $\,\,0\ \cdots\ 0$ & $0$ & $0\ \cdots\ 0$ & $1$ &
      $0\ \cdots\ 0$ & $\,\,\marker$ & &
      $2^j$ & $2^j\le 2^i$ \\ \cline{2-6}
    & $\,\,0\ \cdots\ 0$ & $0$ & $0\ \cdots\ 0$ & $1$ & $0\ \cdots\ 0$ & & &
      $r\xss{1}$ & $r\xss{1}=2^j\bmod k$ \\ \cline{2-6}
    & $\,\,0\ \cdots\ 0$ & $0$ & $0\ \cdots\ 0$ & $1$ & $0\ \cdots\ 0$ & & &
      $r\xss{2}$ & $r\xss{2}=2^j\bmod 2^i$ \\ \cline{2-6}
    & $\,\,0\ \cdots\ 0$ & $0$ & $0\ \cdots\ 0$ & $0$ & $0\ \cdots\ 0$ & & &
      $a$ & $a=0$ \\ \cline{1-7}
  \end{tabular}
\efigure{Initial content on the worktape, created in the course of
  one sweep to the left, followed by one sweep to the right,
  during which the input head travels exactly
  $2\!\cdot\!(\ell\!+\!1)$ positions. The lengths $j,i,\ell$ are
  guessed nondeterministically, and so are the bits in the track
  for~$k$, for the bit positions $0,\ldots,i\!-\!1$, displayed
  here as $\mathquo{0\!/\!1}\!$\@. The condition
  $2^j= 2\!\cdot\!(\ell\!+\!1)$ is not granted, to be verified
  later\@.}{f:worktape}%
\hypertarget{hy:AM.requirement}{}
\bitemize
  \item[$k,2^i$:] guessed, so that $2^i< k< 2^{i+1}$ and
    $k< 2^{\ell}\!$\@. Also~$\ell$, the length of the allocated
    worktape space, is guessed.
  \item[$2^j$:] guessed, so that $2^j\le 2^i$ and
    $2^j= 2\!\cdot\!(\ell\!+\!1)$\@.
  \item[$r\xss{1}$,$r\xss{2}$:] initialized to
    $r\xss{1}= 2^j\bmod k$ \ and \ $r\xss{2}= 2^j\bmod 2^i\!$\@.
    (Clearly, this is equivalent to
    $r\xss{1}= (2\!\cdot\!(\ell\!+\!1))\bmod k$ \ and \
    $r\xss{2}= (2\!\cdot\!(\ell\!+\!1))\bmod 2^i$\@.)
  \item[$a$:] auxiliary track, initialized to $a=0$\@. (Used for
    verification, in Section~\ref{s:verify}\@.)
\eitemize

\medbreak
Before passing to implementation details, let us show that the above
requirements are realistic. More precisely, for each
$1^n\!\in\!\aml$, there exist $k,i,\lbr j,\ell$ satisfying not only
$n\bmod k\ne 0$ \ and \ $n\bmod 2^i=0$, but also additional
requirements imposed by the realtime processing. This can be
achieved by using the following values:
\bequations
  & k= f(n)  \,,\  i= \flor{\log k} \,,\
    j= 2\!+\!\flor{\log(2\!+\!i)}  \,,\  \ell= 2^{j-1}\!-\!1 \,.
\eequations{e:ijkl}
We have to show that $2^i\!< \!k\!< \!2^{i+1}\!$,
\,$k\!< \!2^{\ell}\!$, \,$2^j\!\le 2^i\!$, and
$2^j\!= \!2\!\cdot\!(\ell\!+\!1)$, for each $1^n\!\in\!\aml$\@.

First, since $k=f(n)$ is not a power of~$2$ for $1^n\!\in\!\aml$,
the value $\log k$ is not an integer. This gives
$\flor{\log k}< \log k< \flor{\log k}\!+\!1$, and hence
\bdisplay
  & 2^i= 2^{\flor{\log k}}< 2^{\log k}= k= 2^{\log k}<
    2^{\flor{\log k}+1}= 2^{i+1} .
\edisplay
Second, using the above inequality and~(\ref{e:ijkl}), we get
\bdisplay
  & k< 2^{i+1}= 2^{(2+i)-1}= 2^{2\Uss{\log(2+i)}-1}<
    2^{2\Uss{1+\flor{\log(2+i)}}-1}= 2^{2\Uss{j-1}-1}= 2^{\ell} .
\edisplay
Third, $k= f(n)\ge 17$, by the additional
assumption~(\ref{e:fnn})\@. But then $i= \flor{\log k}\ge 4$\@. Now,
using $2+\flor{\log(2\!+\!i)}\le i$ for each $i\ge 4$, we have
\bdisplay
  & 2^j= 2^{2+\flor{\log(2+i)}}\le 2^i .
\edisplay
Finally, in~(\ref{e:ijkl}), we took $\ell= 2^{j-1}\!-\!1$\@.
Consequently,
\bdisplay
  2^j &=& 2\!\cdot\!(\ell\!+\!1) \,.
\edisplay

In addition, by the use of~(\ref{e:ijkl}), Lemma~\ref{l:fn}, and
the fact that $5< 2\!\cdot\!\log\log n$ for $n\ge 51$
(by~(\ref{e:fnn}), we actually have $n\ge 720720$), we obtain that
\bequations
  \ell &=& 2^{j-1}\!-\!1= 2^{1+\flor{\log(2+i)}}\!-\!1\le
    2^{1+\log(2+i)}\!-\!1= 2\!\cdot\!(2\!+\!i)\!-\!1=
    3+ 2\!\cdot\!i \\
  &=& 3+ 2\!\cdot\!\flor{\log k}\le 3+ 2\!\cdot\!\log k=
    3+ 2\!\cdot\!\log f(n)< 3+ 2\!\cdot\!\log(2\!\cdot\!\log n) \\
  &=& 5+ 2\!\cdot\!\log\log n< 4\!\cdot\!\log\log n \,.
\eequations{e:ell}
This clearly gives $O(\log\log n)$ weak space bound for~\cla\@.

We are now ready to provide implementation details, showing that
the above values can be guessed, incremented, and tested for zero
by single double-sweeps.

\bsubsection{Initialization -- Details}{s:init}
Here we show how the worktape is initialized in a single
double-sweep, with exactly $2\!\cdot\!(\ell\!+\!1)$ steps. We point
out that even though we guess the values $k,i,j,\ell$ that
satisfy~(\ref{e:ijkl}), they are not guessed in the order in which
they appeared in~(\ref{e:ijkl})\@.

\paragraph*{Sweep to the Left:}
Running the input head forward, \cla~moves the worktape head {}from
the right endmarker and rewrites the blank symbols so that the
worktape becomes organized into six parallel tracks,%
\footnote{Formally, the worktape alphabet is
  $\Gamma= \set{\blank,\!\marker}\cup \set{\,[b_1,\ldots,b_6] \st
    b_1,\ldots,b_6\in \set{0,1}}$\@.}
containing binary written $k,2^i,2^j,\lbr r\xss{1},r\xss{2},a$\@.
(See Figure~\ref{f:worktape}\@.) This is done as follows.%
\footnote{Throughout this section, the $t$\xth~bit of a number~$a$
  is denoted by~$\bit{a}{t}$\@. To avoid confusion with other
  notation, we enclose binary strings in quotation marks, e.g.,
  $\mathquo{\,b^i\,}$~represents $i$~replicated copies of the same
  bit $b\in \set{0,1}$ (a~string), while $b^i$~denotes $b$~raised to
  the power of~$i$ (a~number)\@.}
\bitemize
  \item In a loop, running through the bit positions
    $t= 0,\ldots,j\!-\!1$, the machine assigns the $t$\xth~bit in
    the track for~$k$ (that is, the value~$\bit{k}{t}$) by
    guessing, while all bits in the remaining tracks are filled
    with zeros.
  \item At the position $t=j$ (this moment is chosen
    nondeterministically), the bit $\bit{k}{t}$ is guessed while
    the bits in other tracks are set as follows: $\bit{2^i}{t}= 0$,
    \,$\bit{2^j}{t}= \bit{r\xss{1}}{t}= \bit{r\xss{2}}{t}= 1$, and
    $\bit{a}{t}= 0$\@.
  \item In a loop, running through the bit positions
    $t= j\!+\!1,\ldots,i\!-\!1$, the bit $\bit{k}{t}$ is guessed
    and all other bits are set to~$0$\@.
  \item At the position $t=i$ (nondeterministically chosen), we
    set $\bit{k}{t}= \bit{2^i}{t}= 1$, all other bits are set
    to~$0$\@.
  \item In a loop, running through the bit positions
    $t= i\!+\!1,\ldots,\ell\!-\!1$ (for nondeterministically
    chosen~$\ell$), all bits are set to~$0$\@.
\eitemize
As a special case, \cla~may guess $i=j$\@. In this case, the
positions $i$ and~$j$ overlap, the phase running through
$j\!+\!1,\ldots,i\!-\!1$ is skipped, and the bits at the position
$t=i=j$ are set as follows:
$\bit{k}{t}= \bit{2^i}{t}= \bit{2^j}{t}= \bit{r\xss{1}}{t}= 1$ and
$\bit{r\xss{2}}{t}= \bit{a}{t}= 0$\@.

{}From Figure~\ref{f:worktape}, we see that the worktape now
contains the following values:
\bdisplay
  k &=& \mathquo{0^{\ell-i-1}1b_{i-1}\!\cdots b_0}< 2^{i+1}\le 2^{\ell} ,\\
  2^i &=& \mathquo{0^{\ell-i-1}10^i\,}\le k \,,\\
  2^j &=& \mathquo{0^{\ell-j-1}10^j\,}\le 2^i ,\\
  r\xss{1} &=& 2^j= 2^j\bmod k \,,
    & \mbox{ \ provided that $2^i<k$} \,,\\
  r\xss{2} &=& 2^j= 2^j\bmod 2^i ,
    & \mbox{ \ if $i>j$} \,,\\
  r\xss{2} &=& \mathquo{0^{\ell}\,}= 0= 2^i\bmod 2^i= 2^j\bmod 2^i ,
    & \mbox{ \ if $i=j$} \,,\\
  a &=& 0 \,,
\edisplay
for some nondeterministically chosen values $j\le i< \ell$ and
$b_0,\ldots,b_{i-1}\in \set{0,1}$\@.

Note that this initialization ensures automatically all requirements
imposed on $k,i,j,\ell,\lbr r\xss{1},r\xss{2},a$, as
\hyperlink{hy:AM.requirement}{listed} in Section~\ref{s:epsfree}
(hence, no verification needed), except for: \iit{i}~Instead of
$2^i< k< 2^{i+1}\!$, we guarantee only $2^i\le k< 2^{i+1}\!$\@. The
inequality $2^i< k$ will be verified on the way back to the right
worktape endmarker. \iit{ii}~The condition
$2^j= 2\!\cdot\!(\ell\!+\!1)$ is not granted, to be verified later
(described in Section~\ref{s:verify})\@.

\paragraph*{Sweep to the Right:}
Running the input head forward, \cla~moves the worktape head back to
the right endmarker and verifies whether $2^i< k$\@. Since $k$
and~$2^i$ have the most significant bit at the same position~$i$
and, in the track for~$2^i\!$, this is the only bit set to~$1$, it is
sufficient to check whether the track for~$k$ contains at least
two~$1$'s. If $2^i< k$, the machine proceeds to testing membership
of~$1^n$ in~\aml, assuming that the condition
$2^j= 2\!\cdot\!(\ell\!+\!1)$ is valid. If $2^i=k$, the machine
rejects.

\bsubsection{Modular Incrementing and Testing for Zero -- Details}{s:count}
After initialization presented in the previous section, we increment
$r\xss{1},r\xss{2}$, by executing the statements
$r\xss{1}:= (r\xss{1}\!+\!2^j)\bmod k$,
\,$r\xss{2}:= (r\xss{2}\!+\!2^j)\bmod 2^i\!$, and check whether
$r\xss{1}\ne 0$, \,$r\xss{2}= 0$, all this in a single double-sweep,
with $2\!\cdot\!(\ell\!+\!1)$ steps. This is repeated until we get
to the end of the input.

\paragraph*{Sweep to the Left:}
At the beginning, \cla~nondeterministically chooses
\bitemize
  \item between $r\xss{1}:= r\xss{1}\!+\!2^j$ \ or \
    $r\xss{1}:= r\xss{1}\!+\!2^j\!-\!k$, and
  \item between $r\xss{2}:= r\xss{2}\!+\!2^j$ \ or \
    $r\xss{2}:= r\xss{2}\!+\!2^j\!-\!2^i\!$\@.
\eitemize
These two choices are independent, hence, \cla~updates
$r\xss{1},r\xss{2}$ in one of four modes. For the correct
combination, we preserve $r\xss{1}\in \set{0,\ldots,k\!-\!1}$ and
$r\xss{2}\in \set{0,\ldots,2^i\!-\!1}$\@. The respective tracks are
updated simultaneously, traversing {}from the right endmarker to the
first blank, while running the input head forward.

To see details, consider, as an example, implementation of
$r\xss{1}:= r\xss{1}\!+\!2^j\!-\!k$ by a single sweep. With access
to $\bit{r\xss{1}}{t},\bit{2^j}{t},\bit{k}{t}$ at each bit
position $t=0,\ldots,\ell\!-\!1$, we can combine the classical
binary addition and subtraction into a single procedure, keeping a
carry value $c_t\in \set{-1,0,+1}$ in the finite state control
(instead of a carry bit, sufficient for one operation alone)\@.
Starting with $c_t=0$ for $t=0$, the new values of
$\bit{r\xss{1}}{t}$ and~$c_{t+1}$ are determined as follows:
\bdisplay
  c_{t+1} :=
    \flor{(\bit{r\xss{1}}{t}\!+\!\bit{2^j}{t}\!-\!\bit{k}{t}\!+\!c_t)/2}
  &;\ &
  \bit{r\xss{1}}{t} :=
    (\bit{r\xss{1}}{t}\!+\!\bit{2^j}{t}\!-\!\bit{k}{t}\!+\!c_t)\bmod 2 \,.
\edisplay
Finally, if \cla~reaches the first blank symbol~\quo{\blank} on the
worktape with $c_{\ell}=-1$ in the finite state control, it rejects
(wrong guess, resulting in negative~$r\xss{1}$)\@.

The other operations for updating $r\xss{1},r\xss{2}$ are
implemented analogically.

\paragraph*{Sweep to the Right:}
Next, while running the input head forward, \cla~moves the worktape
head back to the right endmarker. During this process the following
four conditions are tested simultaneously:
\bitemize
  \item $r\xss{1}<k$ and $r\xss{2}<2^i\!$\@. If any of these
    conditions is not valid, reject.
  \item $r\xss{1}\qeq 0$ and $r\xss{2}\qeq 0$\@. The outcome of
    these comparisons is kept in the finite state control when we
    get back to the right worktape endmarker.
\eitemize

As an example, to check whether $r\xss{1}<k$, scan the respective
tracks for the first different bit. If
$\bit{r\xss{1}}{t}< \bit{k}{t}$ at some position~$t$, the remaining
bits are ignored. Conversely, if $\bit{r\xss{1}}{t}> \bit{k}{t}$ or
there is no difference at all, reject. The other comparisons,
running simultaneously, are implemented in a similar way.

When the sweep has been completed, i.e., the worktape head is back
at the right endmarker, \cla~starts another double-sweep, to process
the next $2\!\cdot\!(\ell\!+\!1)$ input symbols. If, at this moment,
we have reached the end of the input, \cla~halts. This is done in an
accepting or rejecting state, depending on whether $r\xss{1}\ne 0$
and $r\xss{1}=0$\@. Recall that this information is remembered in
the finite state control.

\medbreak
If \cla~hits the end of the input in the course of execution of this
double-sweep, it rejects; this can happen only if the length of the
input is not an integer multiple of $2\!\cdot\!(\ell\!+\!1)$\@.
Therefore, $n\bmod 2^i\ne 0$, by the reasoning {}from
Section~\ref{s:epsfree}, based on the assumption that
$2\!\cdot\!(\ell\!+\!1)= 2^j$ (not verified here)\@.

\bsubsection{Verifying the Size of the Allocated Space -- Details}{s:verify}
Now we shall verify the condition $\ell= 2^{j-1}\!-\!1$,
equivalent to $2^j= 2\!\cdot\!(\ell\!+\!1)$\@. The verification is
activated after the initialization, presented in
Section~\ref{s:init}, and it runs in parallel with the \quo{main}
procedure counting in $r\xss{1},r\xss{2}$, presented in
Section~\ref{s:count}\@. This causes no problems, since both
routines move both the input and worktape heads in the same way.
We shall keep data on the auxiliary track reserved for~$a$, while
the \quo{main} procedure runs simultaneously on the first five
tracks. (The track for~$2^j$ is shared, however, both routines use
it in a read-only way\@.) When the condition $\ell= 2^{j-1}\!-\!1$
has been confirmed, the verification routine stops. {}From that
moment on, the \quo{main} procedure runs alone. If we find that
$\ell\ne 2^{j-1}\!-\!1$, the computation of the \quo{main}
procedure is aborted and \cla~rejects.

In the first double-sweep, write the bits $0,1,1$ in the auxiliary
track, in that order, starting {}from the right endmarker and moving
to the left. Then traverse across the worktape and return back. This
initializes the track to $a= \mathquo{110}= 6= 2^h\!+\!h$, for
$h=2$\@. If there is no room for storing $\mathquo{110}\!$, reject.
(If $1^n\!\in\!\aml$ and $k,i,j,\ell$ were guessed in accordance
with~(\ref{e:ijkl}), then, by~(\ref{e:fnn}), we have
$\ell> \log k= \log f(n)\ge \log 17> 4$\@.)

Now, in a loop, the machine modifies the auxiliary track as follows.

\paragraph*{Sweep to the Left:}
Moving across the worktape, \cla~increases~$a$, that is,
$a:= a\!+\!1$\@. Now the auxiliary track contains
$a'= 2^h\!+\!(h\!+\!1)$\@. The bit positions for~$2^h$ and $h\!+\!1$
do not overlap, since $\flor{\log(h\!+\!1)}< h$, for each
$h\ge 2$\@.

\paragraph*{Sweep to the Right:}
On the way back, moving across the initial segment of zeros,
\cla~guesses the position $h\!+\!1$ standing in front of the
leftmost~$1$, and sets this bit to $\bit{a}{h+1}= 1$\@. Next,
\cla~verifies whether $\bit{a}{h}= 1$ and clears this bit to
$\bit{a}{h}= 0$\@. (In case of a wrong guess, \cla~rejects\@.) Thus,
the leftmost~$1$ is shifted one position to the left; now the track
contains $a''= 2^{h+1}\!+\!(h\!+\!1)$\@. After that, \cla~returns to
the right endmarker and starts another sweep to the left, for the
new value $h:= h\!+\!1$\@.

However, if there is no initial segment of zeros, i.e., if the
leftmost~$1$ is at the position $\ell\!-\!1$, the machine proceeds
in a different way. At this moment, we clearly have $h= \ell\!-\!1$,
which gives $a'= 2^h\!+\!(h\!+\!1)= 2^{\ell-1}\!+\!\ell$\@. Thus, by
ignoring the leftmost bit $\bit{a}{\ell-1}= 1$, the content of the
auxiliary track becomes equal to~$\ell$, the length of the worktape
written in binary. \cla~proceeds across the worktape to the right
and compares $\ell$ with $2^{j-1}\!-\!1$\@. The comparison is based
on the following two facts: First,
$2^{j-1}\!-\!1= \mathquo{1^{j-1}}\!$, binary written as a string not
containing zeros. Second, the length of this string can be
determined by looking into the track for~$2^j\!$, containing
$\mathquo{0^{\ell-j-1}10^j\,}\!$\@. Thus, in a loop, running through
the bit positions $t= \ell\!-\!1,\ell\!-\!2,\ldots,$ the machine
searches for the first $\bit{2^j}{t}= 1$, checking also if
$\bit{a}{t}= 0$ in the meantime (not taking $\bit{a}{\ell-1}\ne 0$
into account)\@. At the position~$j$, the machine checks if
$\bit{a}{j}= 0$, and then if $\bit{a}{j-1}= 0$\@. After that, in a
loop, for $t= j\!-\!2,\ldots,0$, the machine checks if
$\bit{a}{t}= 1$\@. If the value stored in the auxiliary track passes
the test, we can confirm that $\ell= 2^{j-1}\!-\!1$, the
verification is over. Otherwise, we reject.

\medbreak
If \cla~hits the end of the input before the verification is over,
it rejects, overriding any potential acceptance made by the
\quo{main} procedure. To see that this is enough to fix the problem,
consider how many steps are performed in the course of verification,
i.e., how far we can get along the input tape.

First, the verification is activated after the initialization in
Section~\ref{s:init}, done in a single double-sweep. The
verification routine uses also its own initial double-sweep, to
write down $\mathquo{110}= 2^2\!+\!2$ in the auxiliary track.
Finally, for $h= 2,\ldots,\ell\!-\!1$, the track is updated, {}from
$2^h\!+\!h$ to $2^{h+1}\!+\!(h\!+\!1)$\@. (This includes the last
double-sweep making the final comparisons\@.) Since each
double-sweep across the worktape takes exactly
$2\!\cdot\!(\ell\!+\!1)$ steps, the total number of steps is bounded
by
\bdisplay
  V(n) &=& [\,1+1+(\ell\!-\!2)\,] \times 2\!\cdot\!(\ell\!+\!1) =
    2\!\cdot\!\ell^2 + 2\!\cdot\!\ell \le 4\!\cdot\!\ell^2 \le
    4\!\cdot\!(4\!\cdot\!\log\log n)^2 \\
  &=& 64\!\cdot\!(\log\log n)^2 < n \,,
\edisplay
using~(\ref{e:ell}) and the fact that
$64\!\cdot\!(\log\log n)^2< n$, for each $n\ge 668$\@. But, under
the assumption~(\ref{e:fnn}), we actually have $n\ge 720720$\@.

Summing up, if $1^n\!\in\!\aml$, then either \iit{i}~$k,i,j,\ell$
are guessed correctly, in accordance with~(\ref{e:ijkl}), but
then, by~(\ref{e:fnn}), there is enough time to finish
verification, or \iit{ii}~$k,i,j,\ell$ are not guessed correctly,
possibly violating~(\ref{e:ijkl}), but then a premature rejection
will do no harm. Clearly, for $1^n\!\notin\!\aml$, no kind of
premature rejection can do harm, whatever happens.

\bsubsection{The Finishing Touch}{s:touch}
The above nondeterministic machine never accepts an input
$1^n\!\notin\!\aml$, since there are no positive integers $k$
and~$i$ satisfying $2^i< k< 2^{i+1}\!$, such that $n\bmod k\ne 0$ \
and \ $n\bmod 2^i= 0$\@. Conversely, for each $1^n\!\in\!\aml$, the
machine has at least one accepting computation path, \emph{provided
that} $f(n)\ge 17$, the assumption introduced by~(\ref{e:fnn})\@.

If $1^n\!\in\!\aml$ but $f(n)< 17$, we can still guess $k=f(n)$ with
$2^i= 2^{\flor{\log k}}$ and prove the membership by verifying that
$n\bmod k\ne 0$ \ and \ $n\bmod 2^i= 0$\@. However, the above
algorithm requires also $j,\ell$ satisfying additional
\hyperlink{hy:AM.requirement}{conditions}, not granted for
$k< 17$\@. But, for each fixed constant~$k$, the task of verifying
can be implemented as a finite state automaton counting the length
of the input modulo $k\!\cdot\!2^{\flor{\log k}}\!$\@.

Thus, the updated machine~\cla\ nondeterministically chooses {}from
among \iit{i}~running the procedure presented above,
\iit{ii}~counting the length of the input modulo
$k\!\cdot\!2^{\flor{\log k}}$ in the finite state control, for some%
\footnote{{}From $\set{16,\ldots,1}$, the set of candidates for~$k$,
  we exclude $16,8,4,2$ \,(if $f(n)= 2^m\!$, for some integer
  $m\ge 1$, then $1^n\!\notin\!\aml$) and $15,14,12,10,6,1$
  \,({}from~\cite{BMP94a,BMP94b}, we know that $f(n)= p^m\!$, for
  some prime~$p$ and integer $m\ge 1$)\@. Finally, we exclude~$3$,
  since we do not need an extra cycle counting modulo $3\!\cdot\!2$,
  once we have a cycle counting modulo $9\!\cdot\!8$\@.}
$k\in \set{13,11,\lbr 9,7,5}$\@. The machine starts in an accepting
state, which ensures that $1^0\in\aml$ is accepted.

\bsection{Realtime Nondeterministic PDA -- Theorem~\ref{t:realtime-NPDA}}{s:rtNPDA}
\hypertarget{hy:REI}{}%
The language \rei\ consists of inputs which \emph{are not prefixes}
of the infinite word
\bitemize
  \item $\omega=
      bc_1ac_2\rvs\, bc_2ac_3\rvs \cdots bc_kac_{k+1}\rvs\,
      bc_{k+1}ac_{k+2}\rvs \cdots,$
    where
  \item $c_k=
      eb_0db_{k,0}db_0\rvs\,\, eb_1db_{k,1}db_1\rvs \cdots
      eb_{\flor{\log k}}db_{k,\flor{\log k}}db_{\flor{\log k}}\rvs\, e$\\
    is a counter representation for~$k$, augmented with subcounters,
  \item $b_{k,i}\in \set{0,1}$ is the $i$\xth\ bit in the binary
    representation of~$k$, and $b_i\in \set{0,1}\str$ denotes the
    number~$i$ written in binary, for
    $i\in \set{0,1,\ldots,\flor{\log k}}$\@.
\eitemize

A~realtime nondeterministic PDA accepts a word~$w$ which is not a
prefix of~$\omega$ by guessing and verifying an error, which can be
of the following kind (see also Figure~\ref{f:rtNPDA})\@:
\bfigure
  \setlength{\unitlength}{0.36mm}%
  \begin{picture}(265,100)(0,-10)
  \thicklines
  \put(87,44){\line(-1,4){6}}
  \put(81,68){\line(5,-1){20}}
  \put(63,9){\line(-1,4){6}}
  \put(57,33){\line(5,-1){20}}
  \put(160,45){\line(1,4){6}}
  \put(166,69){\line(-5,-1){20}}
  \put(207,44){\line(-1,4){6}}
  \put(201,68){\line(5,-1){20}}
  \put(83,30){\line(1,0){12}}
  \put(95,30){\line(-1,2){5}}
  \put(75,18){\line(1,0){40}}
  \put(115,18){\line(-1,2){17}}
  \put(168,18){\line(-1,0){40}}
  \put(128,18){\line(1,2){17}}
  \put(203,30){\line(1,0){15}}
  \put(218,30){\line(-2,3){7}}
  \put(195,18){\line(1,0){40}}
  \put(235,18){\line(-1,2){17}}
  \put(80,56){\makebox(0,0)[r]{$\log\log \left\{\mbox{\begin{picture}(0,15)(0,10)\end{picture}}\right.$} }
  \put(245,45){\makebox(0,0)[l]{\phantom{$..$} $\left.\mbox{\begin{picture}(0,50)(0,45)\end{picture}}\right\} \log$} }
  \put(0,0){\makebox(0,0){$b$}}
  \put(5,5){\makebox(0,0){$c_1$}}
  \put(10,10){\makebox(0,0){$a$}}
  \put(15,5){\makebox(0,0){$c_2\rvs$}}
  \put(20,0){\makebox(0,0){$b$}}
  \put(25,8){\makebox(0,0){$c_2$}}
  \put(30,16){\makebox(0,0){$a$}}
  \put(35,8){\makebox(0,0){$c_3\rvs$}}
  \put(40,0){\makebox(0,0){$b$}}
  \put(50,0){\makebox(0,0){$.....$}}
  \put(245,0){\makebox(0,0){$.....$}}
  \put(60,0){\makebox(0,0){$b$}}
  \multiput(64,6)(2,3){3}{\makebox(0,0){.}}
  \put(72,18){\makebox(0,0){$e$}}
  \put(76,24){\makebox(0,0){$b_i$}}
  \put(80,30){\makebox(0,0){$d$}}
  \put(84,36){\makebox(0,0){$b_{k,i}$}}
  \put(88,42){\makebox(0,0){$d$}}
  \put(92,48){\makebox(0,0){$b_i\rvs$}}
  \put(96,54){\makebox(0,0){$e$}}
  \put(100,60){\makebox(0,0){$b_{i+1}$}}
  \put(105,66){\makebox(0,0){$d$}}
  \put(108,72){\makebox(0,0){$b_{k,i+1}$}}
  \multiput(112,78)(2,3){3}{\makebox(0,0){.}}
  \put(121,90){\makebox(0,0){$a$}}
  \multiput(176,6)(-2,3){3}{\makebox(0,0){.}}
  \put(172,18){\makebox(0,0){$e$}}
  \put(170,24){\makebox(0,0){$b_i\rvs$}}
  \put(164,30){\makebox(0,0){$d$}}
  \put(160,36){\makebox(0,0){$b_{k+1,i}$}}
  \put(156,42){\makebox(0,0){$d$}}
  \put(152,48){\makebox(0,0){$b_i$}}
  \put(148,54){\makebox(0,0){$e$}}
  \put(149,60){\makebox(0,0){$b_{i+1}\rvs$}}
  \put(140,66){\makebox(0,0){$d$}}
  \put(136,72){\makebox(0,0){$b_{k+1,i+1}$}}
  \multiput(132,78)(-2,3){3}{\makebox(0,0){.}}
  \put(180,0){\makebox(0,0){$b$}}
  \multiput(184,6)(2,3){3}{\makebox(0,0){.}}
  \put(192,18){\makebox(0,0){$e$}}
  \put(196,24){\makebox(0,0){$b_i$}}
  \put(200,30){\makebox(0,0){$d$}}
  \put(202,36){\makebox(0,0){$b_{k+1,i}$}}
  \put(208,42){\makebox(0,0){$d$}}
  \put(212,48){\makebox(0,0){$b_i\rvs$}}
  \put(216,54){\makebox(0,0){$e$}}
  \put(220,60){\makebox(0,0){$b_{i+1}$}}
  \put(225,66){\makebox(0,0){$d$}}
  \put(228,72){\makebox(0,0){$b_{k+1,i+1}$}}
  \multiput(232,78)(2,3){3}{\makebox(0,0){.}}
  \put(240,90){\makebox(0,0){$a$}}
  \put(63,8){\makebox(0,0)[lt]{$\underbrace{\phantom{bbbbbbbbbbbbbb}}_{c_{k}}$}}
  \put(123,8){\makebox(0,0)[lt]{$\underbrace{\phantom{bbbbbbbbbbbbbb}}_{c_{k+1}\rvs}$}}
  \put(183,8){\makebox(0,0)[lt]{$\underbrace{\phantom{bbbbbbbbbbbbbb}}_{c_{k+1}}$}}
  \end{picture}
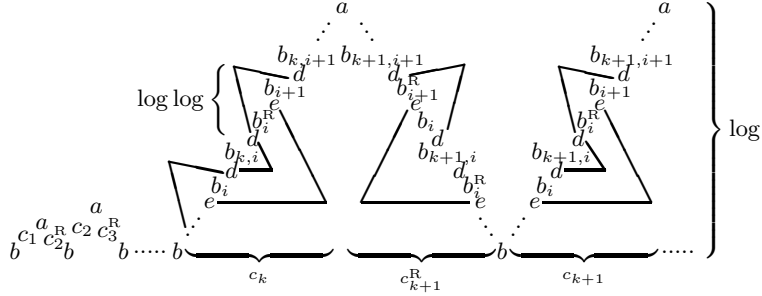
\efigure{The structure of~$\omega$, where each counter
  representation $c_k$ consists of $O(\log k)$ bits, and each of
  these bits is associated with a subcounter of size $O(\log\log k)$
  bits\@.}{f:rtNPDA}

\iit{i}~There is some error in the format, that is, the input $w$ is
not a prefix of any word
$\omega'\in (\,
  b(e\set{0,1}\str d\set{0,1}d\set{0,1}\str)\str e\,
  a(e\set{0,1}\str d\set{0,1}d\set{0,1}\str)\str e\, )\str\!$\@.

\iit{ii}~The input $w$ does not begin with the counter
representation for~$1$, that is, with $bc_1a= b\,e0d1d0e\,a$, \,or
\,$|w|\le |bc_1a|$ but $w$~is not a prefix of~$bc_1a$\@.

\iit{iii}~For some~$k$, the counter representation~$c_k$ does not
begin with the first subcounter, i.e., with $beb_0d= be0d$\@.
Symmetrically, we check whether $c_k\rvs$~does not end by
$db_0\rvs eb= d0eb$\@.

\iit{iv}~For some $k$ and~$i$, the subcounter $b_i$ in~$c_k$ is not
correct, i.e., $c_k$~contains a defective part
\,$db_{i-1}\rvs e\,b_{i'}\,d$ \,or \,$eb_id\xi d\,b_{i'}\rvs \,e$
\,with $i'\ne i$ and $\xi\in \set{0,1}$\@. This can be recognized by
using the pushdown store. Assuming that, within~$c_k$, the $i$\xth\
subcounter is the smallest one with this error, i.e., the
subcounters $b_0,\ldots,b_{i-1}$ are correct, the pushdown space can
be bounded by $O(|b_{i-1}|)\le O(|b_{\flor{\log k}}|)$, even though
$b_{\flor{\log k}}$ is actually not present. Symmetrically, we check
whether $c_k\rvs$~contains a defective~$b_i\rvs$\@.

\iit{v}~All subcounters in~$w$ are correct but, for some~$k$, the
part between two consecutive~$a$'s is of the form
$ac_k\rvs b\,c_{k'}\,a$, with $k'\ne k$\@. This leaves us two
subcases. First, $c_k\rvs$ and~$c_{k'}$ do not agree in the highest
subcounter. This is detected by loading the highest subcounter
in~$c_k\rvs$ (the first one) into the pushdown store and check it
against the highest subcounter in~$c_{k'}$ (the last one)\@. Second,
$k$ and~$k'$ differ in the $i$\xth\ bit, for some~$i$\@. This is
detected by guessing the position of~$b_{k,i}$ in~$c_k\rvs$, pushing
the following subcounter~$b_i$ on the pushdown store, then guessing
the corresponding position in~$c_{k'}$, verifying the subcounter
value there, and checking that $b_{k',i}\ne b_{k,i}$\@.

\iit{vi}~All subcounters in~$w$ are correct but, for some~$k$, the
part between two consecutive~$b$'s is of the form
$bc_{k}a\,c_{k'}\rvs\, b$, with $k'\ne k\!+\!1$\@. Here we have
three subcases. First, the binary written $k$ is not of the
form~$\mathquo{1^{\ell}\,}$ for some $\ell\ge 1$, but $c_k$
and~$c_{k'}\rvs$ do not agree in the highest subcounter. Second, the
binary written $k$ is of the form~$\mathquo{1^{\ell}\,}$ (hence, the
highest subcounter in~$c_k$ is~$b_{\ell-1}$), but the highest
subcounter in~$c_{k'}\rvs$ is not equal to~$b_{\ell}\rvs$\@. Both
these subcases can be recognized similarly as in the previous case.
Third, $k\!+\!1$ and~$k'$ differ in some bit. This is again
recognized similarly as in the previous case: this time we verify
that, for some~$i$, by going {}from $c_k$ to~$c_{k'}\rvs$, either the
$i$\xth\ bit changed but the $(i\!-\!1)$\xst\ bit did not change
{}from $1$ to~$0$, or the $i$\xth\ bit did not change but the
$(i\!-\!1)$\xst\ bit changed {}from $1$ to~$0$ (or $i=0$)\@.

\medbreak
The errors described in the cases \iit{i}--\iit{iii} are detected by
using the finite state control. In the cases \iit{iv}--\iit{vi}, we
only need to store one subcounter. Thus, the size of the pushdown
store can be bounded by $O(|b_{\flor{\log k}}|)\le O(\log\log k)$\@.
If $c_k$~is the smallest counter representation with an error, i.e.,
$c_1,\ldots,c_{k-1}$ are correct, we have indeed $k\!-\!1$ counters
present along the input, and hence $k\!-\!1\le n$\@. Thus, the
pushdown size $O(\log\log n)$ is sufficient to guess and verify the
smallest occurring incorrectness.

\bsection{Realtime Alternating Counter Automaton -- Theorem~\ref{t:realtime-A1CA}}{s:realtime-A1CA}
First, we give a one-way \cla\ for a finite variation of \upower\
which, in one step, either moves the input head or changes the value
in the counter, but not both. Each such step is followed by exactly
one \quo{\eps-step} neither moving the input head nor changing the
counter. (But even \eps-steps depend on whether the counter contains
zero\@.)

The idea is to represent~$k$, the current distance {}from the end of
the input, by parallel processes of an alternating machine~\cla\@.
Each process uses its counter to address only a single bit of~$k$\@.
Here we use two existential states $s_0,s_1$\@: \,\cla~has an
accepting alternating subtree in the configuration $(s_v,k,j)$
---\,corresponding to $s_v\in \set{s_0,s_1}$ with the head $k\ge 0$
positions away {}from the end of the input and a number $j\ge 0$
stored in the counter\,--- if and only if~$b\xss{k,j}$, the
$j$\xth~bit in the binary written~$k$, is equal to~$v$\@. (In~$q_v$,
only \eps-steps are executed\@.)

The computation is based on the fact that $b\xss{k,j}$ depends only
on $b\xss{k-1,j}$, $b\xss{k,j-1}$, and~$b\xss{k-1,j-1}$\@. Namely,
for each $j>0$ and $k>0$, using Boolean notation,
\bdisplay
  b\xss{k,j} &\equiv&
    ( \neg b\xss{k-1,j} \!\!\wedge \!\neg b\xss{k,j-1} \!\!\wedge \!b\xss{k-1,j-1} \!) \!\vee\!
    ( b\xss{k-1,j} \!\!\wedge \!\neg b\xss{k-1,j-1} \!) \!\vee\!
    ( b\xss{k-1,j} \!\!\wedge \!b\xss{k,j-1} \!\!\wedge \!b\xss{k-1,j-1} \!) ,\\
  \neg b\xss{k,j} &\equiv&
    ( \neg b\xss{k-1,j} \!\!\wedge \!\neg b\xss{k-1,j-1} \!) \!\vee\!
    ( \neg b\xss{k-1,j} \!\!\wedge \!b\xss{k,j-1} \!\!\wedge \!b\xss{k-1,j-1} \!) \!\vee\!
    ( b\xss{k-1,j} \!\!\wedge \!\neg b\xss{k,j-1} \!\!\wedge \!b\xss{k-1,j-1} \!) .
\edisplay
For $j=0$ with $k>0$, we have
$b\xss{k,j}\!\!\equiv \!\neg b\xss{k-1,j}$ and, for $k=0$,
\,$b\xss{k,j}\!\!\equiv \!0$\@. Thus, in $s_v\in \set{s_0,s_1}$ and
with the counter not containing zero, \cla~guesses existentially
which of the three clauses leads to $b\xss{k,j}=v$\@. (For
$s_v=s_0$, we have a fourth branch, guessing $k=0$\@. This branch
switches to a state~$\hat{s}$, described later\@.) Next,
\cla~branches universally to verify that all literals in the chosen
clause are valid. This moves the input head and/or decreases the
counter, ending in the state $s_0$ or~$s_1$, depending on whether
the given literal is negated. As an example, in~$s_1$, \,\cla~might
guess, by one \eps-step, that $b\xss{k,j}=1$ because of the clause
$( \neg b\xss{k-1,j} \!\!\wedge \!\neg b\xss{k,j-1} \!\!\wedge
  \!b\xss{k-1,j-1} \!)$\@.
After that, \cla~branches to verify literals: the respective
parallel path \iit{a}~moves the input head by going to~$q_0$,
\iit{b}~decreases the counter by going to~$q_0$, \iit{c}~moves the
input head, executes one \eps-step, and then decreases the counter
by going to~$q_1$\@.

The case of $s_v\in \set{s_0,s_1}$ with the counter containing zero
is similar, utilizing $b\xss{k,0}\!\!\equiv \!\neg b\xss{k-1,0}$\@.
(Also here we have a path to~$\hat{s}$, guessing $k=0$\@.)

Finally, at the end of the input, \cla~enters the state~$\hat{s}$ by
guessing the case of $k=0$\@. This can happen in the state~$s_0$
only, since $b\xss{0,j}\!\!\equiv \!0$\@. Now, in a loop,
\cla~decreases the counter and executes one \eps-step. When the
counter is cleared, \cla~halts and accepts. (If, due to a wrong
guess, \cla~enters $\hat{s}$ with $k>0$, the computation is blocked
in the middle of the input, and hence such path rejects\@.)

Given an input~$a^m\!$, \,\cla~verifies that $m= 2^n$ for some
$n\ge 2$, that is, whether \iit{i}~$b\xss{m,n}=1$,
\iit{ii}~$b\xss{m,j}=0$ for each $j<n$, and \iit{iii}~$b\xss{k,n}=0$
for each $k<m$\@. Thus, \cla~starts with a loop, in which it first
increases the counter and then, by the next \eps-step, it guesses
existentially whether to exit. This chooses some $n\!-\!1\ge 1$\@.
After that, by increasing the counter once more, \cla~branches
universally to verify the conditions \iit{i}--\iit{iii}\@. That is,
\cla~branches to \iit{i}~the state~$s_1$, \iit{ii}~a~universal loop
consisting of one \eps-step followed by one decreasing of the
counter with branching to~$s_0$, \iit{iii}~a~universal loop
consisting of one \eps-step followed by one move of the input head
with branching to~$s_0$\@. (Both these loops halt in accepting
states\@.)

This completes the construction of~\cla\@. For each $a^m$ with
$m= 2^n$ and $n\ge 2$, the accepting alternating subtree is unique,
with all paths moving the input head $2^n$~times and changing the
counter value exactly $2n$~times. Each such step is followed by
exactly one \eps-step. Thus, by making the input head move in every
step, we get a realtime \clap\ accepting $a^{m'}$ with
$m'= 2\!\cdot\!2^n\!+\!4n= 2^{n+1}\!+\!4\!\cdot\!(n\!+\!1)\!-\!4$,
which changes the accepted language {}from
$\upower\!\setminus\!\set{a^1,a^2}$ to~\upowerp\@.

\bsection{A~Trade-Off to Alternation Depth}{s:trade-off}
The machine in the proof of Theorem~\ref{t:realtime-A1CA} uses a
linear number of alternations. However, we can recognize \upower\
with only a logarithmic alternation depth, but using a counter of
linear size. To make this algorithm easier to follow, we construct a
realtime alternating~\cla\ with a counter capable of containing
\emph{also negative integers} and, moreover, in one step, the
counter can be updated by any $\Delta\in \set{-3,\ldots,+3}$,
instead of $\Delta\in \set{-1,0,+1}$\@. \,(\cla~can be easily
modified to meet the standard definition without changing the
language, and hence this extension is not essential\@.)

First, along the input~$a^m\!$, \,\cla~existentially picks a
position~$j_1$, increasing the counter by~$1$ per each input symbol.
Thus, the counter contains $j_1> 0$ and the remaining part of the
input is of length $m_1= m\!-\!j_1$\@. Then \cla~branches
universally:

\iit{i}~In the first branch, \cla~verifies that $j_1=m_1$, i.e.,
that $j_1= m\!-\!j_1$, decreasing the counter by~$1$ per each symbol
until it gets to the end of the input. Thus, this branch is
successful only if $2j_1=m$, i.e., only if $j_1$~is the exact half
of~$m$\@.

\iit{ii}~In the second branch, assume that $j_1=m_1$ and $2j_1=m$
since, for any other values, the outcome is overridden due to the
first branch. Along~$a^{m_1}\!$, \,\cla~existentially picks a new
position~$j_2$, decreasing the counter by~$3$ per each input symbol.
Now the counter contains $j_1\!-\!3j_2< 0$, the rest of the input is
of length $m_2= m_1\!-\!j_2= j_1\!-\!j_2$\@. Then \cla~makes a
universal branching similar to the previous one:

\iit{ii.i}~In the first branch, \cla~verifies that
$-(j_1\!-\!3j_2)= m_2$, i.e., that $-(j_1\!-\!3j_2)= j_1\!-\!j_2$,
increasing the counter by~$1$ per each symbol until it gets to the
end of the input. Thus, this branch is successful only if $2j_2=j_1$\@.

\iit{ii.ii}~In the second branch, assume that $-(j_1\!-\!3j_2)= m_2$
and $2j_2=j_1$, so we start with the counter containing
$j_1\!-\!3j_2= -j_2$ and the rest of the input of length
$m_2= -(j_1\!-\!3j_2)= j_2$\@. Now, along~$a^{m_2}\!$,
\,\cla~existentially picks~$j_3$, increasing the counter by~$3$ per
each input symbol, so the counter contains $-j_2\!+\!3j_3> 0$, with
the rest of the input of length $m_3= m_2\!-\!j_3= j_2\!-\!j_3$\@.
Then \cla~branches universally:

First, in~\iit{ii.ii.i}, \cla~verifies that $(-j_2\!+\!3j_3)= m_3$,
i.e., that $(-j_2\!+\!3j_3)= j_2\!-\!j_3$, decreasing the counter
by~$1$ per each symbol until the end of the input. This branch is
successful only if $2j_3=j_2$\@. Second, in parallel~\iit{ii.ii.ii},
we assume $(-j_2\!+\!3j_3)= m_3$ and $2j_3=j_2$, so \cla~starts with
the counter containing $-j_2\!+\!3j_3= j_3$ and the rest of the
input of length $m_3= -j_2\!+\!3j_3= j_3$\@. Now \cla~proceeds in
the same situation as in~\iit{ii}, with $j_3=m_3$ instead of
$j_1=m_1$ \dots\ \dots

This is repeated until, for some~$i$, there remains $j_i=1$ and a
single input symbol, when \cla~accepts. If $m$~is a power of~$2$, we
have an accepting computation subtree in which
$j_1\!= \!\frac{m}{2}$, $j_2\!= \!\frac{m}{4}$, \dots,
$j_i\!= \!\frac{m}{2^i}$, \dots,
$j\xss{\log m}\!= \!\frac{m}{2^{\log m}}\!= \!1$,
with the counter containing $+j_1,-j_2,+j_3,-j_4,\ldots$ at the
moment of universal choice. This values are unique, leading to a
unique accepting alternating subtree for each accepted input, with a
logarithmic number of alternations. If $m$~is not a power of~$2$,
there is no accepting subtree, since, for some~$j_i$, \,\cla~fails
to find the exact half of the remaining input.

\smallbreak
In a similar way, \upower\ can be recognized with only one
alternation but using a linear pushdown store instead of a counter,
as follows: For the given~$a^m\!$, guessing existentially, the
automaton loads some $w\rvs cw'\in b\set{a,b}\str cb\set{a,b}\str$
into the pushdown store and, branching universally, it verifies that
$|w'|=|a^m|$, $w'=w$, $w_{2^i}=b$ for each $i\ge 0$ (provided that
$2^i\le m$), and that $w_j=a$ at all other positions. (Combining
these conditions, we get that $m= 2^n\!$, for some~$n$\@.) The
verification of $|w'|=|a^m|$ just compares the lengths, by popping
one symbol {}from the pushdown store and by moving one input
position forward, until the symbol~$c$ is popped. But the machine
universally branches at each position in~$w'$, to do the following
two checks:

First, the same letter {}from $\set{a,b}$ that appears $j$~positions
away {}from~$c$ in~$w'$ must also appear $j$~positions away
{}from~$c$ in~$w\rvs\!$\@. (This ensures $w'=w$\@.) To guarantee
equal distance in $w'$ and~$w\rvs\!$, the parallel branch stops the
input head movement until $c$~is popped out, after which the input
head movement is synchronized with popping out again, until the end
of the input is reached.

Second, the same letter {}from $\set{a,b}$ that appears $j$~positions
away {}from~$c$ in~$w'$ must also appear $2j$~positions away {}from~$c$
in~$w\rvs\!$\@. Moreover, in~$w\rvs\!$, the symbol~$a$ must appear
$2j\!+\!1$ positions away. (This ensures all remaining
conditions\@.) Checking this is similar to the previous test but,
after popping~$c$ out, the input head moves only one position
forward per two pushdown symbols popped out.

\bsection{Two-Way Deterministic PDA -- Theorem~\ref{t:two-way-DPDA}}{s:2DPDA}
Here we can follow the same construction as in the proof of
Theorem~\ref{t:realtime-NPDA}\@. (We might even use an easier
version abandoning the reverse written parts\@.) Instead of guessing
the kind of incorrectness, we have to check them one by one in an
appropriate order, to make sure that we find the smallest occurring
incorrectness first:

First, check the counter representation~$c_1$\@. After checking
$c_1,\ldots,c_{k-1}$, the machine checks~$c_k$\@. Namely, we check
that the smallest subcounter is~$b_0$ and that the subcounters are
correctly increasing, until we get a subcounter~$b_i$ equal to the
highest subcounter in~$c_{k-1}$\@. Then check the highest subcounter
in~$c_k$ against~$c_{k-1}$ and that the main counter is increasing
correctly, by going back and forth between $c_{k-1}$ and~$c_k$ for
each bit with the related subcounter in the pushdown store.

To check that subcounters are correctly increasing, just load the
current subcounter to the pushdown store and compare it with the
next one. Even if such comparison fails, we can restore the current
pushdown contents, by going back to the beginning of the tested
counter and using the prefix which, so far, has been identical. To
find a position related to the current subcounter within~$c_{k-1}$,
just load the current subcounter to the pushdown store and compare
it with the subcounters in~$c_{k-1}$, one after another, until the
\quo{proper} one is found. Since we compare without destroying the
pushdown store, we can return back to the original position in~$c_k$
in the same way.

\bsection{Two-Way Quantum Counter Automata -- Theorem~\ref{t:quantum-logspace-counter}}{s:quantum-logspace-counter}
Recently, Yakary{\i}lmaz~\cite{Yak13B} introduced a new programming
technique for 2QCCAs and it was shown that
$\usquare= \set{a^{n^2}\st n\ge 1}$ can be recognized by 2QCCAs for
any error bound by using $O(\sqrt{n})$ space on its counter for all
accepted inputs. Based on this technique, we show that $O(\log n)$
space can also be useful.

\hypertarget{hy:POWER}{}%
2QCFAs can recognize $\power= \set{a^nb^{2^n}\st n\ge 1}$ such that
each $w\in\power$ is accepted with probability~$1$ and each
$w\notin\power$ rejected with a probability arbitrarily close
to~$1$~\cite{YS10B}\@. Let \clp\ be such a 2QCFA, rejecting
$w\notin\power$ with a probability at least~$\frac{8}{9}$
(\hyperlink{hy:2QCFA-for-POWER}{Appendix})\@. An~important property
of~\clp\ is that it reads the input {}from left to right in an
infinite loop and uses $3$ quantum states.

We present a 2QCCA~\clup\ for~\upower\ calling~\clp\ as a subroutine
such that each $a^m$ in \upower\ is accepted with probability~$1$
and with the counter value not exceeding $\log m$\@. \,Each
$a^m\notin\upower$ is rejected with a probability above
$1\!-\!\frac{1}{8^m+1}$\@. The pseudo-code for~\clup\ is given in
Figure~\ref{f:code}\@.
\bfigure
  \bprogram
    \If $m=0$ \Then \Reject \Else \If $m=1$ \Then \Accept \\
    \Loop \+\\
      \For $i:=1$ \To $m$ \Do \+\\
        \Run \clp\ on the input $w'= a^ib^m$ \+\\
          \If \clp\ accepts $w'$ \Then \Exit \For \\
          \If \clp\ rejects $w'$ and $i=m$ \Then \Reject \-\-\\
      \End \For \\
      \Accept with a probability $p$ satisfying $0< p\le \left(\frac{1}{9}\right)^m$ \-\\
    \End \Loop
  \eprogram
\efigure{A~pseudo-code for~\clup, testing whether $w=a^m\in \upower$\@.}{f:code}

Let $w=a^m$ be the input\@. Using the counter, we implement a
for-loop iterated for $i=1,\ldots,m$, in which we simulate~\clp\ on
the input $w'= a^i b^m\!$\@. Clearly, if $a^m$ is in \upower,
\clp~always accepts $w'=a^{\log m}b^m$ and so we exit the for-loop
with the counter containing $i\le \log m< m$\@. After the exit {}from
the for-loop, $a^m$~is accepted with a probability
$p\le (\frac{1}{9})^m$ but, since this process is nested in an outer
infinite loop, \clup~accepts $a^m$ with probability
$\sum_{j=0}^{\infty} p\!\cdot\!(1\!-\!p)^j =1$\@. Moreover, the
counter value (hence, the space complexity) never exceeds
$\log m$\@. Conversely, if $a^m$ is not in \upower, it is rejected
with a probability $p'\ge (\frac{8}{9})^m$ by the for-loop (when
$i=m$)\@. Thus, it is accepted with probability
$(1\!-\!p')\!\cdot\!p$ after the for-loop. Again, since this is
nested in the outer infinite loop, \clup~rejects $a^m$ with
probability
\bdisplay
  && \sum_{j=0}^{\infty} p' \!\cdot \!( 1\!- \!p'\!- \!(1\!-\!p')\!\cdot\!p )^j
    \ge \sum_{j=0}^{\infty} p' \!\cdot \!( 1\!- \!p'\!- \!(\frac{1}{9})^m )^j
    = \frac{p'}{p'+(1/9)^m} \\
  && \phantom{\sum_{j=0}^{\infty}}
    \ge \frac{(8/9)^m}{(8/9)^m+(1/9)^m}
    = 1\!- \!\frac{1}{8^m+1} \,.
\edisplay
This probability can raised closer to~$1$, using $(\frac{1}{c})^m$
with $c>9$ instead of~$(\frac{1}{9})^m\!$\@.

\pdfbookmark{Acknowledgments}{hy.bk:ack}%
\subsection*{Acknowledgments}
We thank Pavol \v{D}uri\v{s} for kindly providing a one-way
alternating automaton for \upower\ using a linear counter, Alexander
Okhotin and Holger Petersen for their answers to our questions, and
the anonymous reviewers for the helpful comments.


\begin{thebibliography}{99}%
%
\bibitem{AM75}
H.~Alt and K.~Mehlhorn.
\newblock A~language over a one symbol alphabet requiring only 
  {$O(\log\log n)$} space.
\newblock {\em SIGACT News}, 7, 31--33, 1975.

\bibitem{AI99}
M.~Amano and K.~Iwama.
\newblock Undecidability on quantum finite automata.
\newblock In {\em Proc.\ ACM Symp.\ Theory of Comput.}, pp.\ 368--75, 1999.

\bibitem{AW02}
A.~Ambainis and J.~Watrous.
\newblock Two--way finite automata with quantum and classical states.
\newblock {\em Theoret.\ Comput.\ Sci.}, 287, 299--311, 2002.

\bibitem{BMP94a}
A.~Bertoni, C.~Mereghetti, and G.~Pighizzini.
\newblock On languages accepted with simultaneous complexity bounds and their
  ranking problem.
\newblock In {\em Proc.\ Math.\ Found.\ Comput.\ Sci.}, {\em Lect.\ Notes
  Comput.\ Sci.},\ 841, pp.\ 245--55. Springer-Verlag, 1994.

\bibitem{BMP94b}
A.~Bertoni, C.~Mereghetti, and G.~Pighizzini.
\newblock An optimal lower bound for nonregular languages.
\newblock {\em Inform.\ Process.\ Lett.}, 50, 289--92, 1994\@.
\newblock (Corr.\ {\em ibid.}, 52, p.~339, 1994\@.)

\bibitem{BMP95}
A.~Bertoni, C.~Mereghetti, and G.~Pighizzini.
\newblock Strong optimal lower bounds for {Turing} machines that accept
  nonregular languages.
\newblock In {\em Proc.\ Math.\ Found.\ Comput.\ Sci.}, {\em Lect.\ Notes
  Comput.\ Sci.},\ 969, pp.\ 309--18. Springer-Verlag, 1995.

\bibitem{CKS81}
{A.\,K.}\ Chandra, {D.\,C.}\ Kozen, and {L.\,J.}\ Stockmeyer.
\newblock Alternation.
\newblock {\em J.~Assoc.\ Comput.\ Mach.}, 28, 114--33, 1981.

\bibitem{Dur13A}
P.~{\v D}uri{\v s}.
\newblock Private communication, October 2013.

\bibitem{DG82B}
P.~{\v D}uri{\v s} and Z.~Galil.
\newblock On reversal-bounded counter machines and on pushdown automata with a
  bound on the size of their pushdown store.
\newblock {\em Inform.\ \& Control}, 54, 217--27, 1982.

\bibitem{DS90}
C.~Dwork and L.~Stockmeyer.
\newblock A~time complexity gap for two-way probabilistic finite-state
  automata.
\newblock {\em SIAM J.\ Comput.}, 19, 1011--23, 1990.

\bibitem{FL75}
{A.\,R.}\ Freedman and {R.\,E.}\ Ladner.
\newblock Space bounds for processing contentless inputs.
\newblock {\em J.~Comput.\ System Sci.}, 11, 118--28, 1975.

\bibitem{Fre81}
R.~Freivalds.
\newblock Probabilistic two-way machines.
\newblock In {\em Proc.\ Math.\ Found.\ Comput.\ Sci.}, {\em Lect.\ Notes
  Comput.\ Sci.},\ 118, pp.\ 33--45. Springer-Verlag, 1981.

\bibitem{Fre83}
R.~Freivalds.
\newblock Space and reversal complexity of probabilistic one-way {Turing}
  machines.
\newblock In {\em Proc.\ Fund.\ Comput.\ Theory}, {\em Lect.\ Notes Comput.\
  Sci.},\ 158, pp.\ 159--70. Springer-Verlag, 1983.

\bibitem{FK94}
R.~Freivalds and M.~Karpinski.
\newblock Lower space bounds for randomized computation.
\newblock In {\em Proc.\ Internat.\ Colloq.\ Automata, Languages, \&
  Programming}, {\em Lect.\ Notes Comput.\ Sci.},\ 820, pp.\ 580--92.
  Springer-Verlag, 1994.

\bibitem{Gab84}
J.~Gabarr{\'o}.
\newblock Pushdown space complexity and related full-{A.F.L.s}.
\newblock In {\em Proc.\ Symp.\ Theoret.\ Aspects Comput.\ Sci.}, {\em Lect.\
  Notes Comput.\ Sci.},\ 166, pp.\ 250--59. Springer-Verlag, 1984.

\bibitem{GR62}
S.~Ginsburg and {H.\,G.}\ Rice.
\newblock Two families of languages related to {ALGOL}.
\newblock {\em J.~Assoc.\ Comput.\ Mach.}, 9, 350--71, 1962.

\bibitem{Iw93}
K.~Iwama.
\newblock {${\rm ASPACE}(o(\log\log n))$} is regular.
\newblock {\em SIAM J.\ Comput.}, 22, 136--46, 1993.

\bibitem{Kan91B}
J.~Ka{\c n}eps.
\newblock Regularity of one-letter languages acceptable by 2-way finite
  probabilistic automata.
\newblock In {\em Proc.\ Fund.\ Comput.\ Theory}, {\em Lect.\ Notes Comput.\
  Sci.},\ 529, pp.\ 287--96. Springer-Verlag, 1991.

\bibitem{KF90}
J.~Ka{\c n}eps and R.~Freivalds.
\newblock Minimal nontrivial space complexity of probabilistic one-way {Turing}
  machines.
\newblock In {\em Proc.\ Math.\ Found.\ Comput.\ Sci.}, {\em Lect.\ Notes
  Comput.\ Sci.},\ 452, pp.\ 355--61. Springer-Verlag, 1990.

\bibitem{KGF97}
J.~Ka{\c n}eps, D.~Geidmanis, and R.~Freivalds.
\newblock Tally languages accepted by {Monte} {Carlo} pushdown automata.
\newblock In {\em Proc.\ RANDOM:\ Randomization \& Approx.\ Tech.\ Comput.\
  Sci.}, {\em Lect.\ Notes Comput.\ Sci.},\ 1269, pp.\ 187--95.
  Springer-Verlag, 1997.

\bibitem{KW97}
A.~Kondacs and J.~Watrous.
\newblock On the power of quantum finite state automata.
\newblock In {\em Proc.\ IEEE Symp.\ Found.\ of Comput.\ Sci.}, pp.\ 66--75,
  1997.

\bibitem{Me08}
C.~Mereghetti.
\newblock Testing the descriptional power of small {Turing} machines on
  nonregular language acceptance.
\newblock {\em Internat.\ J.\ Found.\ Comput.\ Sci.}, 19, 827--43, 2008.

\bibitem{NC10}
{M.\,A.}\ Nielsen and {I.\,L.}\ Chuang.
\newblock {\em Quantum Computation and Quantum Information}.
\newblock Cambridge Univ.\ Press, 10\textsuperscript{th} edit., 2010.

\bibitem{Paz71}
A.~Paz.
\newblock {\em Introduction to Probabilistic Automata}.
\newblock Academic Press, 1971.

\bibitem{Ra63}
{M.\,O.}\ Rabin.
\newblock Probabilistic automata.
\newblock {\em Inform.\ \& Control}, 6, 230--45, 1963.

\bibitem{Re07}
K.~Reinhardt.
\newblock A~tree-height hierarchy of context-free languages.
\newblock {\em Internat.\ J.\ Found.\ Comput.\ Sci.}, 18, 1383--94, 2007.

\bibitem{RY14}
\hypertarget{hy:RY14}{}%
K.~Reinhardt and A.~Yakary{\i}lmaz.
\newblock The minimum amount of useful space: {New} results and new directions.
\newblock In {\em Proc.\ Develop.\ Lang.\ Theory}, {\em Lect.\ Notes Comput.\
  Sci.},\ 8633, pp.\ 315--26. Springer-Verlag, 2014.

\bibitem{RS62}
{J.\,B.}\ Rosser and L.~Schoenfeld.
\newblock Approximate formulae for some functions of prime numbers.
\newblock {\em Illinois J.\ Math.}, 6, 64--94, 1962.

\bibitem{SayY15}
A.~C.~C. Say and A.~Yakary{\i}lmaz.
\newblock Quantum finite automata: {A}~modern introduction.
\newblock In {\em Gruska Festschrift}, {\em Lect.\ Notes Comput.\ Sci.},\ 8808,
  pp.\ 208--22. Springer-Verlag, 2015.

\bibitem{Sz94b}
A.~Szepietowski.
\newblock {\em Turing Machines with Sublogarithmic Space}.
\newblock Springer-Verlag, 1994.

\bibitem{Yak12C}
A.~Yakary{\i}lmaz.
\newblock Superiority of one-way and realtime quantum machines.
\newblock {\em RAIRO Inform.\ Th{\'e}or.\ Appl.}, 46, 615--41, 2012.

\bibitem{Yak13B}
A.~Yakary{\i}lmaz.
\newblock One-counter verifiers for decidable languages.
\newblock In {\em Proc.\ Comput.\ Sci.\ Russia}, {\em Lect.\ Notes Comput.\
  Sci.},\ 7913, pp.\ 366--77. Springer-Verlag, 2013.

\bibitem{Yak13C}
A.~Yakary{\i}lmaz.
\newblock Public qubits versus private coins.
\newblock In {\em Proc.\ Workshop on Quantum and Classical Complexity}, pp.\
  45--60. Univ.\ Latvia Press, 2013.
\newblock ECCC:TR12-130.

\bibitem{YS09B}
A.~Yakary{\i}lmaz and A.~C.~C. Say.
\newblock Efficient probability amplification in two-way quantum finite
  automata.
\newblock {\em Theoret.\ Comput.\ Sci.}, 410, 1932--41, 2009.

\bibitem{YS10B}
A.~Yakary{\i}lmaz and A.~C.~C. Say.
\newblock Succinctness of two-way probabilistic and quantum finite automata.
\newblock {\em Discrete Math.\ \& Theoret.\ Comput.\ Sci.}, 12, 19--40, 2010.

\bibitem{YS13B}
A.~Yakary{\i}lmaz and A.~C.~C. Say.
\newblock Tight bounds for the space complexity of nonregular language
  recognition by real-time machines.
\newblock {\em Internat.\ J.\ Found.\ Comput.\ Sci.}, 24, 1243--53, 2013.
%
\end{thebibliography}
\pdfbookmark{References}{hy.bk:ref}%

\pdfbookmark{Appendix -- the 2QCFA P for POWER}{hy.bk:2QCFA-for-POWER}%
\hypertarget{hy:2QCFA-for-POWER}{}%
\section*{Appendix. The 2QCFA \clp\ for \power}
The description of~\clp\ is as follows. Let $w\in \set{a,b}\str$ be
an input. We can assume the input of the form $a^mb^n\!$, where
$m>0$ and $n>0$\@. Otherwise, \clp~rejects.

The quantum register has three states:
$\ket{q_1},\ket{q_2},\ket{q_3}$\@. \,\clp~encodes $2^m$ and~$n$ into
amplitudes of $\ket{q_2}$ and~$\ket{q_3}$ and compares them by
subtracting. The resulting amplitude is zero if and only if the
amplitudes are equal. Based on this, the input is rejected. Since we
use only rational numbers, we can bound a nonzero rejecting
probability {}from below, with zero probability only for the members.
With a carefully tuned accepting probability, the members are only
accepted while the nonmembers are rejected with a probability that
is sufficiently high. Since this gap is achieved only with a small
probability, we run the procedure in an infinite loop.

In each iteration (round), the input is read {}from left to right in a
realtime mood. At the beginning of each round, the quantum state is
set to $\ket{\psi_0}= (1\ 0\ 0)\trs\!$\@. We keep the unconditional
quantum state until we read the left endmarker. Then
$\cle_{\ccent}= \set{E_{\ccent,1},E_{\ccent,2}}$ is applied to the
quantum register, i.e.,
\bdisplay
  E_{\ccent,1} = \frac{1}{2} \cmatrix{rrr}{
    1 & 0 & 0 \\[-0.5ex]
    1 & 0 & 0 \\[-0.5ex]
    0 & 0 & 2 }
  &\mbox{ \ and \ }&
  E_{\ccent,2} = \frac{1}{2} \cmatrix{rrr}{
    1 & 0 & 0 \\[-0.5ex]
    1 & 0 & 0 \\[-0.5ex]
    0 & 2 & 0 } ,
\edisplay
where \iit{i}~the current round continues if the outcome
\quo{$\ccent$,$1$} is observed, and \iit{ii}~the current round is
terminated without any decision if the outcome \quo{$\ccent$,$2$} is
observed. Before reading~$a$'s, the quantum state is
\bdisplay
  \ket{\widetilde{\psi_0}} &=&
  \frac{1}{2} \cdot\!\cmatrix{c}{
    1 \\[-0.5ex]
    1 \\[-0.5ex]
    0 } .
\edisplay
For each~$a$, \,$\cle_{a}= \set{E_{a,1},E_{a,2}}$ is applied to the
quantum register, i.e.,
\bdisplay
  E_{a,1} = \frac{1}{2} \cmatrix{rrr}{
    1 & 0 & 0 \\[-0.5ex]
    0 & 2 & 0 \\[-0.5ex]
    0 & 0 & 2 }
  &\mbox{ \ and \ }&
  E_{a,2} = \frac{1}{2} \cmatrix{rrr}{
    1 & 0 & 0 \\[-0.5ex]
    1 & 0 & 0 \\[-0.5ex]
    1 & 0 & 0 } ,
\edisplay
where \iit{i}~the current round continues if the outcome
\quo{$a$,$1$} is observed, and \iit{ii}~the current round is
terminated without any decision if the outcome \quo{$a$,$2$} is
observed. Before reading~$b$'s, the quantum state is
\bdisplay
  \ket{\widetilde{\psi_m}} &=&
  \left( \frac{1}{2} \right)^{m+1} \!\cdot\!\cmatrix{c}{
    1 \\[-0.5ex]
    2^m \\[-0.5ex]
    0 } .
\edisplay
For each~$b$, \,$\cle_{b}= \set{E_{b,1},E_{b,2},E_{b,3}}$ is applied
to the quantum register, i.e.,
\bdisplay
  E_{b,1} = \frac{1}{2} \cmatrix{rrr}{
    1 & 0 & 0 \\[-0.5ex]
    0 & 1 & 0 \\[-0.5ex]
    1 & 0 & 1 } ,\
  E_{b,2} = \frac{1}{2} \cmatrix{rrr}{
    1 & 0 & -\!1 \\[-0.5ex]
    1 & 0 & 0 \\[-0.5ex]
    0 & 1 & 1 } ,
  &\mbox{ \ and \ }&
  E_{b,3} = \frac{1}{2} \cmatrix{rrr}{
    0 & 1 & -\!1 \\[-0.5ex]
    0 & 1 & 0 \\[-0.5ex]
    0 & 0 & 0 } ,
\edisplay
where \iit{i}~the current round continues if the outcome
\quo{$b$,$1$} is observed, and \iit{ii}~the current round is
terminated without any decision if the outcome \quo{$b$,$2$}
or~\quo{$b$,$3$} is observed. Before reading the right endmarker,
the quantum state is
\bdisplay
  \ket{\widetilde{\psi_{|w|}}} &=&
  \left( \frac{1}{2} \right)^{m+n+1} \!\cdot\!\cmatrix{c}{
    1 \\[-0.5ex]
    2^m \\[-0.5ex]
    n } .
\edisplay
When reading the right endmarker,
$\cle_{\dollar}=
  \set{E_{\dollar,1},E_{\dollar,2},E_{\dollar,3},E_{\dollar,4}}$
is applied to the quantum register, i.e.
\bdisplay
  E_{\dollar,1} = \frac{1}{4} \cmatrix{rrr}{
    1 & 0 & 0 \\[-0.5ex]
    0 & 0 & 0 \\[-0.5ex]
    0 & 0 & 0 } ,\
  E_{\dollar,2} = \frac{1}{4} \cmatrix{rrr}{
    0 & 0 & 0 \\[-0.5ex]
    0 & 2 & -\!2 \\[-0.5ex]
    0 & 2 & -\!2 } ,\
  E_{\dollar,3} = \frac{1}{4} \cmatrix{rrr}{
    0 & 2 & 2 \\[-0.5ex]
    0 & 2 & 2 \\[-0.5ex]
    3 & 0 & 0 } ,
  &\mbox{ \ and \ }&
  E_{\dollar,4} = \frac{1}{4} \cmatrix{rrr}{
    2 & 0 & 0 \\[-0.5ex]
    1 & 0 & 0 \\[-0.5ex]
    0 & 0 & 0 } ,
\edisplay
where the actions based on the measurement outcomes are as follows:
\bitemize
  \item the input is accepted if the outcome \quo{$\dollar$,$1$} is
    observed,
  \item the input is rejected if the outcome \quo{$\dollar$,$2$} is
    observed, and
  \item the current round is terminated without any decision,
    otherwise.
\eitemize
Thus, if the outcome \quo{$\dollar$,$1$} is observed, the quantum
state is
\bdisplay
  \ket{\widetilde{\psi_{|w|+1}}} &=&
  \left( \frac{1}{2} \right)^{m+n+3} \!\cdot\!\cmatrix{c}{
    1 \\[-0.5ex]
    0 \\[-0.5ex]
    0 } .
\edisplay
That is, in a single round, the input is always accepted with
probability $(\frac{1}{4})^{m+n+3}\!$\@. If the outcome
\quo{$\dollar$,$2$} is observed, then the quantum state is
\bdisplay
  \ket{\widetilde{\psi_{|w|+1}}} &=&
  \left( \frac{1}{2} \right)^{m+n+3} \!\cdot\!\cmatrix{c}{
    0 \\[-0.5ex]
    2(2^m\!-\!n) \\[-0.5ex]
    2(2^m\!-\!n) } .
\edisplay
That is, in a single round, the input will always be rejected with a
probability
$(\frac{1}{4})^{m+n+3}\!\cdot\!8\!\cdot\!(2^m\!-\!n)^2\!$, which is
\bitemize
  \item zero for any member and
  \item at least $8$ times greater than the accepting probability
    for any nonmember.
\eitemize
Thus, we can conclude that \clp\ accepts any member with
probability~$1$ and rejects any nonmember with a probability at
least~$\frac{8}{9}$\@.
\end{document}